\documentclass[aps,prl,reprint,twocolumn,superscriptaddress,longbibliography,nofootinbib,floatfix,showpacs]{revtex4-1}
\usepackage{epsfig}
\usepackage{amsmath,amssymb,amsfonts}
\usepackage{mathrsfs}
\usepackage{bbm}
\usepackage{slashed}
\usepackage{graphicx}
\usepackage{verbatim}
\usepackage[T1]{fontenc}
\usepackage[utf8]{inputenc}
\usepackage[colorlinks]{hyperref}
\usepackage{ifthen}
\usepackage{forloop}
\usepackage[english]{babel}
\usepackage{mathbbol}
\usepackage[usenames]{color}

\definecolor{darkgreen}{rgb}{0.2,0.6,0}
\definecolor{lightblue}{rgb}{0,0.5,0.8}
\definecolor{lightred}{rgb}{0.8,0.2,0.2}
\definecolor{darkorange}{rgb}{1,0.549,0}

\newcommand{\cG}{{\mathcal G}}
\newcommand{\cM}{{\mathcal M}}
\newcommand{\cR}{{\mathcal R}}

\newcommand{\br}{{\bf r}}
\newcommand{\bq}{{\bf q}}

\newcommand{\p}{{\partial}}

\graphicspath{{figures/}}

\newcommand{\be}{\begin{equation}}
\newcommand{\ee}{\end{equation}}
\newcommand{\ba}{\begin{eqnarray}}
\newcommand{\ea}{\end{eqnarray}}

\begin{document}

\title{Resolving Spacetime Singularities within Asymptotic Safety}
\author{Lando Bosma}
\email[]{l.bosma@science.ru.nl}
\author{Benjamin Knorr}
\email[]{b.knorr@science.ru.nl}
\author{Frank Saueressig}
\email[]{f.saueressig@science.ru.nl}
\affiliation{
Institute for Mathematics, Astrophysics and Particle Physics (IMAPP),\\
Radboud University Nijmegen, Heyendaalseweg 135, 6525 AJ Nijmegen, The Netherlands
}

\begin{abstract}
A key incentive of quantum gravity is the removal of spacetime singularities plaguing the classical theory. We compute the non-perturbative momentum-dependence of a specific structure function within the gravitational asymptotic safety program which encodes the quantum corrections to the graviton propagator for momenta above the Planck scale. The resulting quantum corrected Newtonian potential approaches a constant negative value as the distance between the two point masses goes to zero, thereby removing the classical singularity. The generic nature of the underlying mechanism suggests that it will remain operative in the context of black hole and cosmic singularities.
\end{abstract}

\maketitle
General relativity provides a well-established theory for gravity from sub-millimeter up to cosmic scales \cite{Will:2014kxa}. It has been extremely successful in predicting phenomena like the bending of light by a gravitational field, the gravitational redshift of photons or the existence of gravitational waves. Another striking feature of general relativity is that its solutions rather generically contain specific points where the curvature of spacetime diverges, so-called singularities \cite{Hawking:1969sw}. Well-known examples are the curvature singularities of classical black holes. This feature is often paraphrased as "general relativity predicting its own breakdown" \cite{Hawking:1979ig} and provides one of the central motivations for the search of a more complete theory of gravity, commonly referred to as ``quantum gravity''. Conversely, any candidate for such a theory should explain the fate of these spacetime singularities. In Loop Quantum Gravity, black hole singularities may be removed by quantum geometry effects \cite{Ashtekar:2005cj,Ashtekar:2005qt,Modesto:2006mx}, also see \cite{Barrau:2018rts} for a recent review. Similarly, the fuzzball proposal \cite{Lunin:2001jy,Mathur:2005zp} provides a mechanism for obtaining regular black holes in the framework of string theory. 
 For the gravitational asymptotic safety program \cite{Niedermaier:2006wt,Reuter:2012id,Percacci:2017fkn,Eichhorn:2018yfc,Reuter:2019byg}, the method of renormalization group improvement suggests that black hole singularities may be removed by quantum effects \cite{Bonanno:2000ep,Saueressig:2015xua}, also see \cite{Pawlowski:2018swz,Platania:2019kyx} for the current status and further references. The present work takes a key step towards understanding the fate of spacetime singularities within asymptotic safety, based on a first principles computation. Our main finding is displayed in Fig.\ \ref{Fig.mainresult}, showing that the short-distance divergence in the spin 2-channel of Newton's gravitational potential is resolved by quantum gravity effects. 

\begin{figure}[h]
	\includegraphics[width= 0.45\columnwidth]{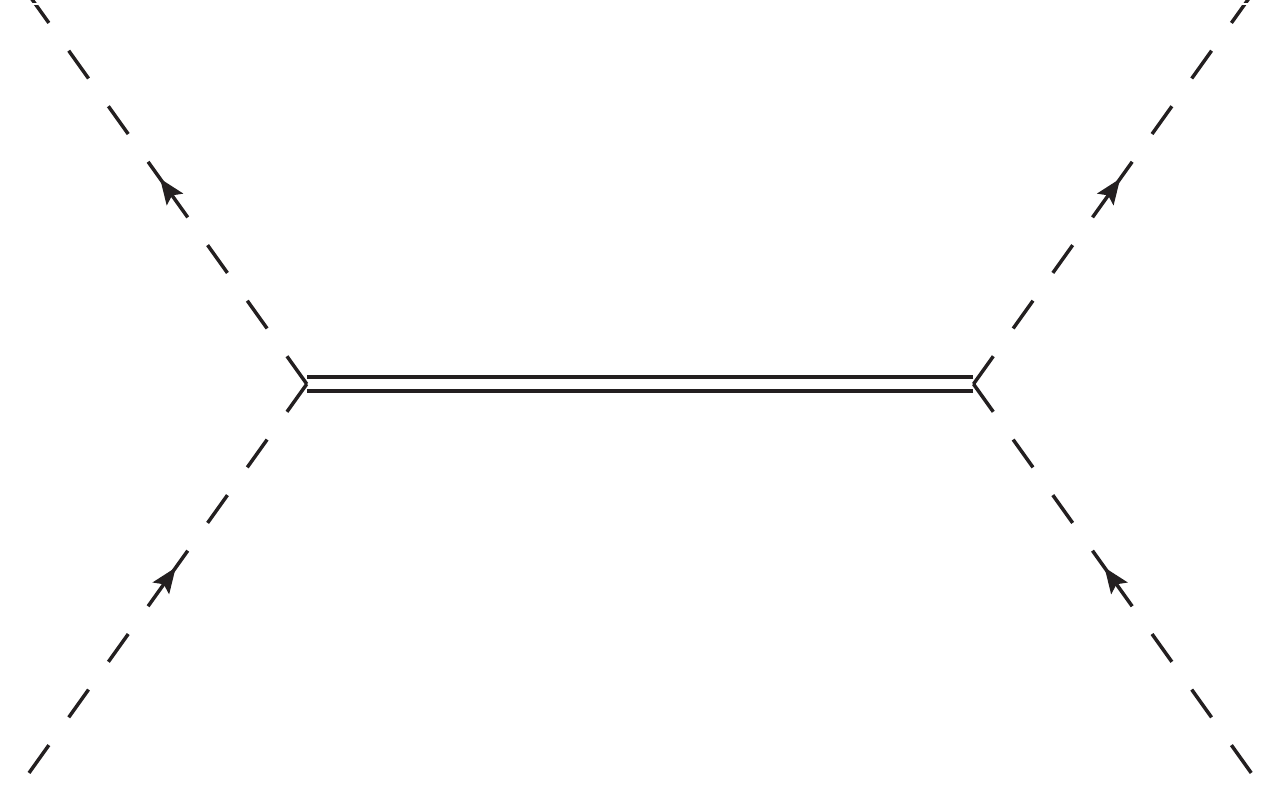}
	\caption{Tree-level amplitude describing the interaction of two scalars of mass $m_1$ and $m_2$ (dashed lines) due to the exchange of a graviton (double line).}  
	\label{Fig2}
\end{figure}

In order to exhibit this effect we follow the path taken in the effective field theory treatment of quantum gravity \cite{Donoghue:1993eb,Donoghue:2017pgk} and construct the gravitational potential $V(r)$ arising from the one-graviton exchange between two scalar fields with masses $m_1$ and $m_2$ minimally coupled to gravity. Taking the static limit where the two scalars have infinite mass one has \cite{Akhundov:2006gh,Donoghue:2017pgk}
\be\label{Vrdef}
V(r) = - \frac{1}{2m_1} \frac{1}{2m_2} \int \frac{d^3 \bq}{(2\pi)^3} \, e^{i \bq \cdot \br}  \cM \, .
\ee
Denoting Newton's coupling by $G$, the scattering amplitude associated with the Feynman diagram Fig.\ \ref{Fig2}, evaluated in general relativity is $\cM = 16 \pi G \, m_1^2 \, m_2^2/|{\bq}|^2$, evaluated for the non-relativistic limit of the propagator $q^\alpha = (0, \bq)$. 
Evaluating the Fourier integral one recovers the classical Newtonian gravitational potential $V_c(r) = - G m_1 m_2/r$.

In the following we will focus on the contribution of the transverse-traceless (spin 2) mode to $\cM$.\footnote{The complete analysis should also include the quantum corrections to the gravitational propagator for the spin-0 mode. This requires adding the structure function $R f_k(\Delta)R$ to the ansatz for eq. \eqref{Gammaans} encoding the non-trivial momentum dependence of the scalar propagator. This analysis is beyond the scope of this work and will be presented elsewhere \cite{us2}. We expect that this will give rise to similar corrections as the ones entering \eqref{fsprop}. } Performing the tensor contractions and taking the static limit one finds 
\be\label{Mamplitude}
\cM^{\rm TT} = \frac{64 \pi G}{3} \, m_1^2 \, m_2^2 \, \, \cG^{\rm TT}(\bq^2) \, ,
\ee
where $\cG^{\rm TT}$ is the scalar part of the spin-2 propagator carrying the momentum dependence, and is obtained from the full propagator by a contraction with the transverse-traceless projector. For the Einstein-Hilbert action $\cG^{\rm TT} = 1/q^2$ so that the resulting potential $V_c^{\rm TT} \propto 1/r$ diverges as $r \rightarrow 0$. In this sense, the non-relativistic limit already includes many of the essential features related to the curvature singularity encountered in black hole physics.

Treating gravity as an effective field theory allows to compute the leading (long-distance) quantum corrections to $V_c(r)$ using perturbation theory \cite{Donoghue:1993eb,Donoghue:2017pgk}. These corrections do not resolve the divergence in $V_c(r)$ though. Eq.\ \eqref{Mamplitude} then suggests to compute the \emph{non-perturbative} propagator  $\cG^{\rm TT}(\bq^2)$ and to investigate the resulting short-distance behavior of the quantum corrected potential $V^{\rm TT}_q(r)$. In this work we perform such a computation within the gravitational asymptotic safety program.
%

{\it Structure functions for Gravity.} A canonical tool for computing properties of a quantum field theory beyond the realm of perturbation theory is the Wetterich equation \cite{Wetterich:1992yh,Morris:1993qb}
\be\label{FRGE}
k \p_k \Gamma_k = \frac{1}{2} {\rm Tr}\left[ \left(\Gamma_k^{(2)} + \cR_k \right)^{-1}  \, k \p_k \cR_k \right] \, ,
\ee
governing the change of the effective average action $\Gamma_k$ when quantum fluctuations around the momentum scale $k$ are integrated out. Here $\Gamma_k^{(2)}$ denotes the second functional derivative of $\Gamma_k$ with respect to the fluctuation field, $\cR_k$ is a suitable regulator function which provides a mass-term for fluctuations with momenta $p^2 < k^2$ and vanishes for $p^2 \gg k^2$, and the trace contains a sum over fluctuation fields as well as an integral over loop momentum. By now, the Wetterich equation has proven its merits in statistical physics \cite{Berges:2002ew}, non-equilibrium physics \cite{Berges:2012ty} and gauge theories \cite{Gies:2006wv}. Starting from the pioneering work \cite{Reuter:1996cp}, which introduced the functional renormalization group in the context of gravity, there is now solid evidence supporting the existence of a non-trivial renormalization group fixed point for four-dimensional gravity \cite{Reuter:2012id,Codello:2007bd,Benedetti:2009rx,Benedetti:2009gn,Manrique:2010am,Becker:2014qya,Gies:2015tca,Ohta:2015fcu,Gies:2016con,Dietz:2016gzg,Falls:2016wsa,Denz:2016qks,Knorr:2017fus,Knorr:2017mhu,Gonzalez-Martin:2017gza,Falls:2017lst,Christiansen:2017bsy,Alkofer:2018fxj,Alkofer:2018baq,Gubitosi:2018gsl,Falls:2018ylp,Eichhorn:2018ydy,Eichhorn:2018akn} and many gravity-matter systems potentially including the standard model of particle physics \cite{Eichhorn:2018yfc,Dona:2013qba,Meibohm:2016mkp,Eichhorn:2016esv,Eichhorn:2016vvy,Christiansen:2017cxa,Christiansen:2017gtg,Eichhorn:2017ylw,Eichhorn:2017lry,Eichhorn:2017muy,Eichhorn:2018nda,Pawlowski:2018ixd,Eichhorn:2018whv}. In particular, the full momentum dependence of the gravitational propagators starting from $\Gamma_k$ given by the Einstein-Hilbert action has been studied in \cite{Christiansen:2014raa,Christiansen:2015rva}. Assuming that this fixed point controls the short distance behavior of the gravitational interaction for lengths $\ell$ smaller than the Planck length $\ell_\text{Pl} \approx 10^{-35}m$ would put gravity in the class of non-perturbatively renormalizable quantum field theories along the lines of Weinberg's asymptotic safety scenario \cite{Weinberg:1976xy}.

A virtue of Wetterich's equation \eqref{FRGE} is that one can extract non-perturbative information about a given quantum theory by making a suitable ansatz for $\Gamma_k$ and studying the flow of the effective average action on the corresponding subspace. A suitable ansatz capturing the momentum-dependence of the gravitational propagator involves scale-dependent structure functions acting on curvature tensors. In this work, we will focus on the non-trivial momentum dependence of the spin-2 propagator captured by
\be\label{Gammaans}
\Gamma_k^{\rm grav} \!\! = \!\! \tfrac{1}{16 \pi G_k} \!\! \int d^4x \sqrt{g} \Big[  - R + 2 \Lambda_k   + C_{\mu\nu\rho\sigma} \, W_k(\Delta) \,  C^{\mu\nu\rho\sigma} \Big] \, . 
\ee
Here $R$, $C_{\mu\nu\rho\sigma}$, and $\Delta \equiv - g^{\mu\nu}D_\mu D_\nu$ denote the Ricci scalar, Weyl tensor, and Laplacian constructed from the spacetime metric $g_{\mu\nu}$, respectively. Furthermore, the ansatz contains a scale-dependent Newton's coupling $G_k$, cosmological constant $\Lambda_k$ as well as the scale-dependent structure function $W_k(\Delta)$. Expanding \eqref{Gammaans} in fluctuations around flat space and restricting the result to the transverse-traceless sector yields the graviton propagator
\be\label{fsprop}
\cG^{\rm TT}(q^2) = \left(q^2  + 2 \, (q^2)^2 \, W_k(q^2) \right)^{-1} \, . 
\ee
 Thus the structure function $W_k(q^2)$ captures non-trivial corrections to the  graviton propagator. The Einstein-Hilbert result is recovered by setting $W_k(q^2) = 0$.

The scale-dependence of $G_k$, $\Lambda_k$ and $W_k(q^2)$ can be obtained by supplementing the ansatz \eqref{Gammaans} by suitable gauge-fixing and ghost terms, substituting the resulting expression into Wetterich's equation \eqref{FRGE} and projecting the trace on the subspace spanned by the ansatz. The calculation of the flow equations for Newton's coupling and the cosmological constant  takes into account the full fluctuation spectrum. Owed to the formidable complexity of the computation, the flow of  $W_k$ is evaluated in the conformally reduced setting \cite{Reuter:2008wj,Reuter:2008qx} where the right-hand side of the flow equation retains the fluctuations of the conformal mode only. In this case, the spacetime metric $g_{\mu\nu}$ is taken to be of the form
$
g_{\mu\nu} = \left(1  + \tfrac{1}{4} h \right) \hat{g}_{\mu\nu}, 
$
where $h$ is the fluctuation field and $\hat{g}_{\mu\nu}$ is a fixed, but arbitrary reference metric. From analogous computations in the framework of $f(R)$-gravity \cite{Demmel:2012ub,Demmel:2014sga,Demmel:2015oqa,Dietz:2016gzg,Dietz:2015owa}, it is expected that the resulting qualitative behavior of the structure function matches the one obtained from including all metric fluctuations.

Our goal is to find a self-consistent flow equation retaining the full information on the functional form of $W_k(\Delta)$, i.e., without making approximations related to the momentum dependence. We achieve this goal by combining two computational techniques tailored to the two classes of curvature terms appearing in the trace evaluation. Terms containing less than two powers of a (potentially contracted) Riemann tensor are evaluated using Mellin transform techniques \cite{Codello:2008vh} together with the non-local heat-kernel results \cite{Codello:2012kq}. Terms containing two powers of the Weyl tensor are evaluated using flat space momentum-space techniques.\footnote{The technical details on how the exact momentum dependence is retained and the conceptual relation between the structure function $W_k(\Delta)$ and the momentum-dependent anomalous dimensions $\eta(p^2)$ studied within the vertex expansion of $\Gamma_k$ \cite{Christiansen:2014raa,Christiansen:2015rva,Meibohm:2015twa,Eichhorn:2018akn} are provided in \cite{Knorr:CQG}.} The resulting flow equations are conveniently expressed in terms of the dimensionless, scale-dependent couplings
\be
g \equiv G_k \, k^{2} \, , \; \lambda \equiv \Lambda_k \, k^{-2} \, , \; w(q^2) \equiv k^{-2} \, W_k(\Delta/k^2) \, . 
\ee 
Neglecting the contribution of the structure function, the flow in the Einstein-Hilbert sector is governed by \cite{Reuter:1996cp, Reuter:2001ag}
\be\label{EHflow}
\begin{split}
	k\p_k \lambda = & \left(\eta_N-2\right) \lambda \\
	& + \frac{g}{2\pi} \left( 10\Phi^1_2(-2\lambda) - 8\Phi^1_2(0) - 5 \eta_N \tilde{\Phi}^1_2(-2\lambda) \right) \, , \\
	k \p_k g = & \left(2 + \eta_N \right) g \, ,
\end{split}
\ee
with the anomalous dimension of Newton's coupling $\eta_N \equiv k \p_k \ln G_k$ being given by
\be
\eta_N = \tfrac{\frac{g}{3\pi} \left[ 5 \Phi^1_1(-2\lambda) - 18 \Phi^2_2(-2\lambda) - 4 \Phi^1_1(0) - 6\Phi^2_2(0) \right]}{1+\frac{g}{6\pi} \left[ 5 \tilde\Phi^1_1(-2\lambda) - 18 \tilde\Phi^2_2(-2\lambda) \right]} \, .
\ee
The threshold functions 
\be\label{EHthreshold}
\begin{split}
\Phi^p_n(\mu) = & \frac{1}{\Gamma(n)} \int_0^\infty dz z^{n-1} \frac{R(z) - z R'(z)}{(z + R(z) + \mu)^p} \, , \\ 
\tilde{\Phi}^p_n(\mu) = & \frac{1}{\Gamma(n)} \int_0^\infty dz z^{n-1} \frac{R(z)}{(z + R(z) + \mu)^p} \, , 
\end{split}
\ee
contain the dimensionless profile function $R(z)$ related to $\cR_k$ \cite{Reuter:1996cp}.

The flow of $w(q^2)$ is treated in the conformally reduced approximation. This leads to the linear integro-differential equation
\begin{widetext}
\begin{equation}\label{wflow}
\begin{aligned}
 k \p_k w(q^2) &= (2+\eta_N) w(q^2) + 2q^2 w'(q^2) + \frac{g}{24\pi} \int_0^{\frac{1}{4}} \text{d}u \, (1-4u)^{\frac{3}{2}} \frac{(2-\eta_N) R(u q^2) - 2 u q^2 R'(u q^2)}{u q^2 + R(u q^2) -\frac{4}{3}\lambda } \\
 &\qquad + \frac{16g}{3\pi^2} \int_0^\infty \text{d}p \int_{-1}^1 \text{d}x \, p^3 \sqrt{1-x^2} \frac{(2-\eta_N) R(p^2) - 2 p^2 R'(p^2)}{(p^2+R(p^2)-\frac{4}{3}\lambda)^2} \Bigg[ \frac{1}{8} \left( w(p^2+2p q x+q^2) - w(q^2) \right) \\
 &\qquad\quad+ \tfrac{2q^4 + 4(q^2-p^2)(p q x) + p^2 q^2(7-6x^2)}{16(p^2+2p q x)^2} \left( w(p^2+2p q x+q^2) - w(q^2) \right) + \tfrac{3 p^4 - 2q^4 + 22p^2 (p q x) -5p^2 q^2(1-6x^2)}{16(p^2+2p q x)} w'(q^2) \Bigg] \, .
\end{aligned}
\end{equation}
\end{widetext}
Here the primes denote derivatives with respect to the argument and $q$ is the dimensionless external momentum. The inhomogeneous term appearing in the first line originates from the Einstein-Hilbert sector. Thus the quantum fluctuations from classical gravity will induce a non-trivial structure function $w(q^2)$ unless $g=0$. While the denominators in the square brackets suggest that the equation could contain collinear divergences, expanding the integrand at these points shows that this is not the case. All potential poles are canceled by zeros of the numerator.

Our primary interest is in non-trivial fixed point solutions\footnote{It is straightforward to see that the system has a trivial fixed point $g_*=\lambda_*=w_*(q^2)=0$.} $(g_\ast, \lambda_\ast, w_\ast(q^2))$ of eqs.\ \eqref{EHflow} and \eqref{wflow} where, by definition, the couplings become independent of $k$. Eq.\ \eqref{EHflow} entails that at such a fixed point $\eta_N = -2$. Substituting this value into eq.\ \eqref{wflow} one finds that the resulting fixed point equation is invariant under a constant shift of $w$. Thus the equation contains one free parameter which will be denoted by $w_\infty$. This freedom constitutes an artefact of the conformally reduced approximation and does not persist once fluctuations of transverse-traceless modes are included. In order to obtain the global form of the structure function $w_*(q^2)$ we first perform an asymptotic expansion of eq.\ \eqref{wflow} at infinite momentum. This establishes the leading order behavior
	\begin{equation}\label{eq.asymptotics}
	w_*(q^2) \underset{q\to\infty}{\sim} w_\infty + \frac{\rho}{q^2} + \ldots \, .
	\end{equation}
	The parameter $w_\infty$ fixes the value of $w_*(q^2)$ at asymptotically large momenta, and $\rho$ is a regulator-dependent positive number.
	
The system of fixed point equations can then be further analyzed numerically. For this purpose we resort to the regulator
$
R(z) = e^{-\alpha z} .
$
Importantly, this regulator is smooth and leads to a rapid convergence when the threshold integrals are evaluated numerically. All numerical values and illustrations are obtained with $\alpha = 1 $ and we checked that all results are robust with respect to changing $\alpha$.
 
Since the Einstein-Hilbert sector is independent of $w$, its fixed point structure can be analyzed before solving eq.\ \eqref{wflow}. It permits a non-Gaussian fixed point at $g_* = 0.374, \lambda_* = 0.285$. This fixed point acts as an ultraviolet attractor for the renormalization group flow in the $g$-$\lambda$-plane.
 
\begin{figure}[h]
	\includegraphics[width= 0.8\columnwidth]{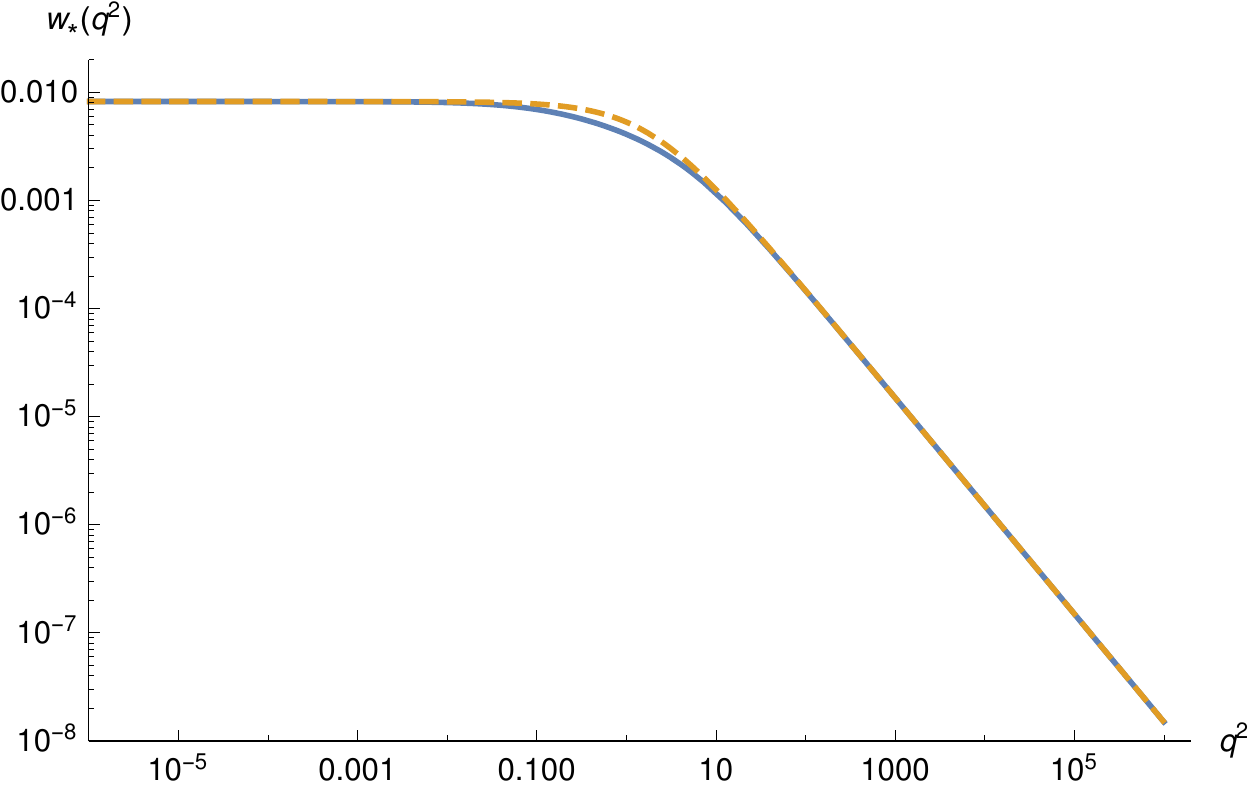}
	\caption{Fixed function $w_*(q^2)$ for $w_\infty = 0$ (orange, solid line).  The parameterization \eqref{wfit} is superimposed as dashed line.}  
	\label{Fig.prop}
\end{figure}

\begin{figure}[t]
	\includegraphics[width= 0.8\columnwidth]{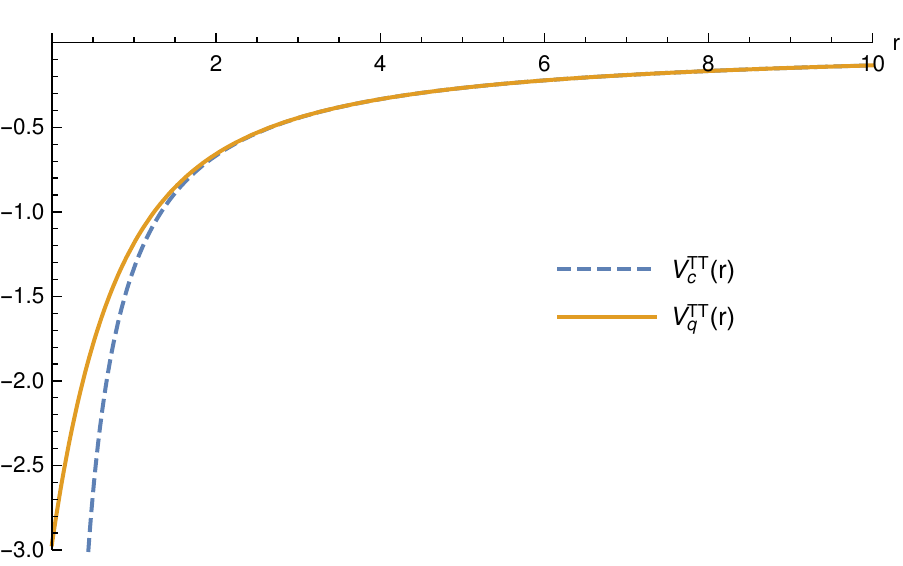}
	\caption{Comparison of $V_c^{\rm TT}(r)$ (blue, dashed line) and $V_q^{\rm TT}(r)$ obtained from the quantum-corrected propagator with $w_\infty = 0.1$ and $m_1 = m_2 = G = 1$. For $w_\infty > 0$ corrections to the high-momentum behavior resolve the divergence in the classical case, making the quantum-corrected potential $V_q(r)$ finite as $r \rightarrow 0$.}  
	\label{Fig.mainresult}
\end{figure}

 The global solution for $w_*(q^2)$ is then obtained through pseudo-spectral methods \cite{Borchardt:2015rxa, Borchardt:2016pif} using rational Chebyshev functions as a basis set \cite{Boyd:ratCheb,Boyd:ChebyFourier}. This leads to the solution shown in Fig.\ \ref{Fig.prop}.
This result has a number of remarkable properties. Firstly, the solution is globally well-defined and unique up to the constant $w_\infty$. The structure function interpolates between a constant for low momenta and the asymptotic behavior \eqref{eq.asymptotics} for large momenta. The crossover occurs for the dimensionless momentum $q^2 \approx 1$. Secondly, the solution is positive definite for all values $w_\infty > 0$. This entails that the flat-space propagator \eqref{fsprop} has a single first-order pole at $q^2 = 0$. In particular there are no additional poles for $q^2 >0$.\footnote{While it would be interesting to extend the analysis of the pole-structure to the complex plane, this is beyond the scope of this letter.} Thirdly, the propagator only grows polynomially for asymptotically large momenta, indicating that the resulting theory is actually local. For the asymptotic parameter, we find $\rho \approx 0.0149$.
  
Remarkably, the numerical solution can be parameterized with very high precision by
\be\label{wfit}
w_*^{\rm fit}(q^2) \approx \frac{\rho}{\frac{\rho}{\kappa} + q^2} + w_\infty \, , \qquad \kappa \approx 0.00817 \, . 
\ee
We expect that this analytic approximation will be very useful when analyzing properties of the quantum theory in the future.

The stability analysis should not be extended to structure functions related to propagators. Conceptually, such structure functions ought to be considered as a part of a momentum-dependent wave function renormalisation. The critical properties of these structure functions are thus related to a momentum-dependent generalisation of the anomalous dimension rather than to the critical exponents, see \cite{Christiansen:2014raa}.\footnote{We thank J.\ M.\ Pawlowski for discussion on this point.}

{\it Quantum corrected Newtonian potential.} As a first application of our computation, we calculate the quantum corrected Newtonian potential $V_q^{\rm TT}(r)$ by evaluating eq.\ \eqref{Vrdef} for the quantum corrected flat-space propagator. This requires reintroducing a scale in $w(q^2)$. The analysis \cite{Reuter:2001ag,Reuter:2004nx,Gubitosi:2018gsl,Saueressig:2019fmx} then indicates that $k^2$ should be identified with the observed value for Newton's coupling $G^{-1}$ which implies that the transition displayed in Fig.\ \ref{Fig.prop} occurs at the Planck scale. While this procedure may miss non-analytic contributions to $w(q^2)$ arising from integrating the renormalization group flow  to the infrared, it is clear that the fixed point will control the short-distance behavior, so that we can make reliable statements about this quantum gravity dominated regime.

The central result is illustrated in Fig.\ \ref{Fig.mainresult}. For distances larger than the Planck scale the classical and quantum Newtonian potentials essentially coincide. For $w_\infty > 0$ the quantum corrections to the propagator remove the short-distance singularity in the classical Newton potential, however, such that $\lim_{r \rightarrow 0} V_q^{\rm TT}(r)$ is actually finite. 
This entails that the gravitational binding energy is bounded by $E_{\rm Binding}^{\rm TT} = - V^{\rm TT}_q(r)|_{r=0}$. The parameterization \eqref{wfit} then allows to compute this value
\be
E_{\rm Binding}^{\rm TT} = \frac{2}{3w_\infty} \frac{1+\frac{\rho}{\kappa \sqrt{x_+ x_-}}}{\sqrt{x_+}+\sqrt{x_-}} \, G m_1 m_2  \, , 
\ee
where
\begin{equation}
\begin{aligned}
 x_\pm = \frac{1}{4w_\infty \kappa} \bigg[ &\kappa +2(w_\infty + \kappa) \rho \\
& \pm \sqrt{(\kappa+2(w_\infty+\kappa)\rho)^2 - 8w_\infty \kappa \rho} \bigg] \, .
\end{aligned}
\end{equation}
The classical analysis \cite{Giacchini:2016xns} establishes that the resolution of the $r =0$-divergence is actually  \emph{independent of the precise form of the structure function}. Generically, any positive structure function will lead to complex mass-poles which ensure that $E_{\rm Binding}^{\rm TT}$ is finite. Deriving the finiteness of $V_c^{\rm TT}$ from a first-principles computation is highly non-trivial and constitutes a major test of the underlying quantum theory. We expect that this will have drastic consequences for our understanding of spacetime singularities also in more general cases. In particuler, it was argued in Ref.\ \cite{Calmet:2017qqa} that complex mass poles in the gravitational propagator could be associated with extended objects which could screen spacetime singularities from being probed by physical processes thus leading to singularity avoidance in the context of black hole physics.

{\it Conclusions.} This work constitutes a major step towards computing the quantum-corrected propagators in asymptotic safety. The non-perturbative short-distance corrections to the Newtonian potential shown in Fig.\ \ref{Fig.mainresult} outline the path for resolving the spacetime singularities plaguing classical gravity. This result differs from the perturbative treatment of gravity as an effective field theory \cite{Donoghue:1993eb,Donoghue:2017pgk} since the propagator underlying $V_q^{\rm TT}(r)$ is manifestly non-perturbative. In principle, the modifications in the Newtonian potential can be tested experimentally (see \cite{Edholm:2016hbt} for a related discussion), even though probing Newton's law on Planckian scales is far beyond current experimental possibilities. 

Naturally, our findings bear a close connection to the ghost-free, non-local gravity program \cite{Biswas:2005qr,Modesto:2011kw,Biswas:2011ar,Modesto:2017sdr} and to non-commutative geometry \cite{Connes:1996gi,vandenDungen:2012ky,1414300} where structure functions of the type \eqref{Gammaans} play a key role. In non-local, ghost-free gravity they constitute an input, defining the fundamental action while the non-commutative geometry approach generates these terms through the non-local heat-kernel. In both cases, the structure function exhibits an exponential fall-off at momentum scales above the non-locality scale. In \cite{Kurkov:2013kfa,Alkofer:2014raa} this has been paraphrased as ``high-energy bosons do not propagate''. The result of our first-principles computation differs qualitatively from these constructions as the quantum corrected propagator arising from \eqref{wflow} grows as $q^4$ for large momenta. This suggests that these approaches are in a different universality class than our (microscopically) manifestly local theory.

\bigskip

\acknowledgments

We thank F.\ Filthaut, S.\ Lippoldt and C.\ Ripken for interesting discussions.
This research is supported by the Netherlands Organisation for Scientific Research (NWO) within
the Foundation for Fundamental Research on Matter (FOM) grant 13VP12.


%


\begin{thebibliography}{91}%
	\makeatletter
	\providecommand \@ifxundefined [1]{%
		\@ifx{#1\undefined}
	}%
	\providecommand \@ifnum [1]{%
		\ifnum #1\expandafter \@firstoftwo
		\else \expandafter \@secondoftwo
		\fi
	}%
	\providecommand \@ifx [1]{%
		\ifx #1\expandafter \@firstoftwo
		\else \expandafter \@secondoftwo
		\fi
	}%
	\providecommand \natexlab [1]{#1}%
	\providecommand \enquote  [1]{``#1''}%
	\providecommand \bibnamefont  [1]{#1}%
	\providecommand \bibfnamefont [1]{#1}%
	\providecommand \citenamefont [1]{#1}%
	\providecommand \href@noop [0]{\@secondoftwo}%
	\providecommand \href [0]{\begingroup \@sanitize@url \@href}%
	\providecommand \@href[1]{\@@startlink{#1}\@@href}%
	\providecommand \@@href[1]{\endgroup#1\@@endlink}%
	\providecommand \@sanitize@url [0]{\catcode `\\12\catcode `\$12\catcode
		`\&12\catcode `\#12\catcode `\^12\catcode `\_12\catcode `\%12\relax}%
	\providecommand \@@startlink[1]{}%
	\providecommand \@@endlink[0]{}%
	\providecommand \url  [0]{\begingroup\@sanitize@url \@url }%
	\providecommand \@url [1]{\endgroup\@href {#1}{\urlprefix }}%
	\providecommand \urlprefix  [0]{URL }%
	\providecommand \Eprint [0]{\href }%
	\providecommand \doibase [0]{http://dx.doi.org/}%
	\providecommand \selectlanguage [0]{\@gobble}%
	\providecommand \bibinfo  [0]{\@secondoftwo}%
	\providecommand \bibfield  [0]{\@secondoftwo}%
	\providecommand \translation [1]{[#1]}%
	\providecommand \BibitemOpen [0]{}%
	\providecommand \bibitemStop [0]{}%
	\providecommand \bibitemNoStop [0]{.\EOS\space}%
	\providecommand \EOS [0]{\spacefactor3000\relax}%
	\providecommand \BibitemShut  [1]{\csname bibitem#1\endcsname}%
	\let\auto@bib@innerbib\@empty
	\bibitem [{\citenamefont {Will}(2014)}]{Will:2014kxa}%
	\BibitemOpen
	\bibfield  {author} {\bibinfo {author} {\bibfnamefont {Clifford~M.}\
			\bibnamefont {Will}},\ }\bibfield  {title} {\enquote {\bibinfo {title} {{The
					Confrontation between General Relativity and Experiment}},}\ }\href {\doibase
		10.12942/lrr-2014-4} {\bibfield  {journal} {\bibinfo  {journal} {Living Rev.
				Rel.}\ }\textbf {\bibinfo {volume} {17}},\ \bibinfo {pages} {4} (\bibinfo
		{year} {2014})},\ \Eprint {http://arxiv.org/abs/1403.7377} {arXiv:1403.7377
		[gr-qc]} \BibitemShut {NoStop}%
	\bibitem [{\citenamefont {Hawking}\ and\ \citenamefont
		{Penrose}(1970)}]{Hawking:1969sw}%
	\BibitemOpen
	\bibfield  {author} {\bibinfo {author} {\bibfnamefont {S.~W.}\ \bibnamefont
			{Hawking}}\ and\ \bibinfo {author} {\bibfnamefont {R.}~\bibnamefont
			{Penrose}},\ }\bibfield  {title} {\enquote {\bibinfo {title} {{The
					Singularities of gravitational collapse and cosmology}},}\ }\href {\doibase
		10.1098/rspa.1970.0021} {\bibfield  {journal} {\bibinfo  {journal} {Proc.
				Roy. Soc. Lond.}\ }\textbf {\bibinfo {volume} {A314}},\ \bibinfo {pages}
		{529--548} (\bibinfo {year} {1970})}\BibitemShut {NoStop}%
	\bibitem [{\citenamefont {Hawking}\ and\ \citenamefont
		{Israel}(1979)}]{Hawking:1979ig}%
	\BibitemOpen
	\bibfield  {author} {\bibinfo {author} {\bibfnamefont {S.~W.}\ \bibnamefont
			{Hawking}}\ and\ \bibinfo {author} {\bibfnamefont {W.~(Eds.)}\ \bibnamefont
			{Israel}},\ }\href
	{http://www.cambridge.org/us/knowledge/isbn/item1131443/?site_locale=en_US}
	{\emph {\bibinfo {title} {{General Relativity - an Einstein Centenary
					Survey}}}}\ (\bibinfo  {publisher} {Univ. Pr.},\ \bibinfo {address}
	{Cambridge, UK},\ \bibinfo {year} {1979})\BibitemShut {NoStop}%
	\bibitem [{\citenamefont {Ashtekar}\ and\ \citenamefont
		{Bojowald}(2005)}]{Ashtekar:2005cj}%
	\BibitemOpen
	\bibfield  {author} {\bibinfo {author} {\bibfnamefont {Abhay}\ \bibnamefont
			{Ashtekar}}\ and\ \bibinfo {author} {\bibfnamefont {Martin}\ \bibnamefont
			{Bojowald}},\ }\bibfield  {title} {\enquote {\bibinfo {title} {{Black hole
					evaporation: A Paradigm}},}\ }\href {\doibase 10.1088/0264-9381/22/16/014}
	{\bibfield  {journal} {\bibinfo  {journal} {Class. Quant. Grav.}\ }\textbf
		{\bibinfo {volume} {22}},\ \bibinfo {pages} {3349--3362} (\bibinfo {year}
		{2005})},\ \Eprint {http://arxiv.org/abs/gr-qc/0504029} {arXiv:gr-qc/0504029
		[gr-qc]} \BibitemShut {NoStop}%
	\bibitem [{\citenamefont {Ashtekar}\ and\ \citenamefont
		{Bojowald}(2006)}]{Ashtekar:2005qt}%
	\BibitemOpen
	\bibfield  {author} {\bibinfo {author} {\bibfnamefont {Abhay}\ \bibnamefont
			{Ashtekar}}\ and\ \bibinfo {author} {\bibfnamefont {Martin}\ \bibnamefont
			{Bojowald}},\ }\bibfield  {title} {\enquote {\bibinfo {title} {{Quantum
					geometry and the Schwarzschild singularity}},}\ }\href {\doibase
		10.1088/0264-9381/23/2/008} {\bibfield  {journal} {\bibinfo  {journal}
			{Class. Quant. Grav.}\ }\textbf {\bibinfo {volume} {23}},\ \bibinfo {pages}
		{391--411} (\bibinfo {year} {2006})},\ \Eprint
	{http://arxiv.org/abs/gr-qc/0509075} {arXiv:gr-qc/0509075 [gr-qc]}
	\BibitemShut {NoStop}%
	\bibitem [{\citenamefont {Modesto}(2008)}]{Modesto:2006mx}%
	\BibitemOpen
	\bibfield  {author} {\bibinfo {author} {\bibfnamefont {Leonardo}\
			\bibnamefont {Modesto}},\ }\bibfield  {title} {\enquote {\bibinfo {title}
			{{Black hole interior from loop quantum gravity}},}\ }\href {\doibase
		10.1155/2008/459290} {\bibfield  {journal} {\bibinfo  {journal} {Adv. High
				Energy Phys.}\ }\textbf {\bibinfo {volume} {2008}},\ \bibinfo {pages}
		{459290} (\bibinfo {year} {2008})},\ \Eprint
	{http://arxiv.org/abs/gr-qc/0611043} {arXiv:gr-qc/0611043 [gr-qc]}
	\BibitemShut {NoStop}%
	\bibitem [{\citenamefont {Barrau}\ \emph {et~al.}(2018)\citenamefont {Barrau},
		\citenamefont {Martineau},\ and\ \citenamefont {Moulin}}]{Barrau:2018rts}%
	\BibitemOpen
	\bibfield  {author} {\bibinfo {author} {\bibfnamefont {Aurelien}\
			\bibnamefont {Barrau}}, \bibinfo {author} {\bibfnamefont {Killian}\
			\bibnamefont {Martineau}}, \ and\ \bibinfo {author} {\bibfnamefont {Flora}\
			\bibnamefont {Moulin}},\ }\bibfield  {title} {\enquote {\bibinfo {title} {{A
					status report on the phenomenology of black holes in loop quantum gravity:
					Evaporation, tunneling to white holes, dark matter and gravitational
					waves}},}\ }\href {\doibase 10.3390/universe4100102} {\bibfield  {journal}
		{\bibinfo  {journal} {Universe}\ }\textbf {\bibinfo {volume} {4}},\ \bibinfo
		{pages} {102} (\bibinfo {year} {2018})},\ \Eprint
	{http://arxiv.org/abs/1808.08857} {arXiv:1808.08857 [gr-qc]} \BibitemShut
	{NoStop}%
	\bibitem [{\citenamefont {Lunin}\ and\ \citenamefont
		{Mathur}(2002)}]{Lunin:2001jy}%
	\BibitemOpen
	\bibfield  {author} {\bibinfo {author} {\bibfnamefont {Oleg}\ \bibnamefont
			{Lunin}}\ and\ \bibinfo {author} {\bibfnamefont {Samir~D.}\ \bibnamefont
			{Mathur}},\ }\bibfield  {title} {\enquote {\bibinfo {title} {{AdS / CFT
					duality and the black hole information paradox}},}\ }\href {\doibase
		10.1016/S0550-3213(01)00620-4} {\bibfield  {journal} {\bibinfo  {journal}
			{Nucl. Phys.}\ }\textbf {\bibinfo {volume} {B623}},\ \bibinfo {pages}
		{342--394} (\bibinfo {year} {2002})},\ \Eprint
	{http://arxiv.org/abs/hep-th/0109154} {arXiv:hep-th/0109154 [hep-th]}
	\BibitemShut {NoStop}%
	\bibitem [{\citenamefont {Mathur}(2005)}]{Mathur:2005zp}%
	\BibitemOpen
	\bibfield  {author} {\bibinfo {author} {\bibfnamefont {Samir~D.}\
			\bibnamefont {Mathur}},\ }\bibfield  {title} {\enquote {\bibinfo {title}
			{{The Fuzzball proposal for black holes: An Elementary review}},}\ }\bibfield
	{booktitle} {\emph {\bibinfo {booktitle} {{The quantum structure of
					space-time and the geometric nature of fundamental interactions. Proceedings,
					4th Meeting, RTN2004, Kolymbari, Crete, Greece, September 5-10, 2004}}},\
	}\href {\doibase 10.1002/prop.200410203} {\bibfield  {journal} {\bibinfo
			{journal} {Fortsch. Phys.}\ }\textbf {\bibinfo {volume} {53}},\ \bibinfo
		{pages} {793--827} (\bibinfo {year} {2005})},\ \Eprint
	{http://arxiv.org/abs/hep-th/0502050} {arXiv:hep-th/0502050 [hep-th]}
	\BibitemShut {NoStop}%
	\bibitem [{\citenamefont {Niedermaier}\ and\ \citenamefont
		{Reuter}(2006)}]{Niedermaier:2006wt}%
	\BibitemOpen
	\bibfield  {author} {\bibinfo {author} {\bibfnamefont {Max}\ \bibnamefont
			{Niedermaier}}\ and\ \bibinfo {author} {\bibfnamefont {Martin}\ \bibnamefont
			{Reuter}},\ }\bibfield  {title} {\enquote {\bibinfo {title} {{The Asymptotic
					Safety Scenario in Quantum Gravity}},}\ }\href {\doibase 10.12942/lrr-2006-5}
	{\bibfield  {journal} {\bibinfo  {journal} {Living Rev.Rel.}\ }\textbf
		{\bibinfo {volume} {9}},\ \bibinfo {pages} {5--173} (\bibinfo {year}
		{2006})}\BibitemShut {NoStop}%
	\bibitem [{\citenamefont {Reuter}\ and\ \citenamefont
		{Saueressig}(2012)}]{Reuter:2012id}%
	\BibitemOpen
	\bibfield  {author} {\bibinfo {author} {\bibfnamefont {Martin}\ \bibnamefont
			{Reuter}}\ and\ \bibinfo {author} {\bibfnamefont {Frank}\ \bibnamefont
			{Saueressig}},\ }\bibfield  {title} {\enquote {\bibinfo {title} {{Quantum
					Einstein Gravity}},}\ }\href {\doibase 10.1088/1367-2630/14/5/055022}
	{\bibfield  {journal} {\bibinfo  {journal} {New J.Phys.}\ }\textbf {\bibinfo
			{volume} {14}},\ \bibinfo {pages} {055022} (\bibinfo {year} {2012})},\
	\Eprint {http://arxiv.org/abs/1202.2274} {arXiv:1202.2274 [hep-th]}
	\BibitemShut {NoStop}%
	\bibitem [{\citenamefont {Percacci}(2017)}]{Percacci:2017fkn}%
	\BibitemOpen
	\bibfield  {author} {\bibinfo {author} {\bibfnamefont {Robert}\ \bibnamefont
			{Percacci}},\ }\href {\doibase 10.1142/10369} {\emph {\bibinfo {title} {{An
					Introduction to Covariant Quantum Gravity and Asymptotic Safety}}}},\
	\bibinfo {series} {100 Years of General Relativity}, Vol.~\bibinfo {volume}
	{3}\ (\bibinfo  {publisher} {World Scientific},\ \bibinfo {year}
	{2017})\BibitemShut {NoStop}%
	\bibitem [{\citenamefont {Eichhorn}(2018)}]{Eichhorn:2018yfc}%
	\BibitemOpen
	\bibfield  {author} {\bibinfo {author} {\bibfnamefont {Astrid}\ \bibnamefont
			{Eichhorn}},\ }\bibfield  {title} {\enquote {\bibinfo {title} {{An
					asymptotically safe guide to quantum gravity and matter}},}\ }\href@noop {}
	{\  (\bibinfo {year} {2018})},\ \Eprint {http://arxiv.org/abs/1810.07615}
	{arXiv:1810.07615 [hep-th]} \BibitemShut {NoStop}%
	\bibitem [{\citenamefont {Reuter}\ and\ \citenamefont
		{Saueressig}(2019)}]{Reuter:2019byg}%
	\BibitemOpen
	\bibfield  {author} {\bibinfo {author} {\bibfnamefont {Martin}\ \bibnamefont
			{Reuter}}\ and\ \bibinfo {author} {\bibfnamefont {Frank}\ \bibnamefont
			{Saueressig}},\ }\href
	{https://www.cambridge.org/academic/subjects/physics/theoretical-physics-and-mathematical-physics/quantum-gravity-and-functional-renormalization-group-road-towards-asymptotic-safety?format=HB&isbn=9781107107328}
	{\emph {\bibinfo {title} {{Quantum Gravity and the Functional Renormalization
					Group}}}}\ (\bibinfo  {publisher} {Cambridge University Press},\ \bibinfo
	{year} {2019})\BibitemShut {NoStop}%
	\bibitem [{\citenamefont {Bonanno}\ and\ \citenamefont
		{Reuter}(2000)}]{Bonanno:2000ep}%
	\BibitemOpen
	\bibfield  {author} {\bibinfo {author} {\bibfnamefont {Alfio}\ \bibnamefont
			{Bonanno}}\ and\ \bibinfo {author} {\bibfnamefont {Martin}\ \bibnamefont
			{Reuter}},\ }\bibfield  {title} {\enquote {\bibinfo {title} {{Renormalization
					group improved black hole space-times}},}\ }\href {\doibase
		10.1103/PhysRevD.62.043008} {\bibfield  {journal} {\bibinfo  {journal} {Phys.
				Rev.}\ }\textbf {\bibinfo {volume} {D62}},\ \bibinfo {pages} {043008}
		(\bibinfo {year} {2000})},\ \Eprint {http://arxiv.org/abs/hep-th/0002196}
	{arXiv:hep-th/0002196 [hep-th]} \BibitemShut {NoStop}%
	\bibitem [{\citenamefont {Saueressig}\ \emph {et~al.}(2016)\citenamefont
		{Saueressig}, \citenamefont {Alkofer}, \citenamefont {D'Odorico},\ and\
		\citenamefont {Vidotto}}]{Saueressig:2015xua}%
	\BibitemOpen
	\bibfield  {author} {\bibinfo {author} {\bibfnamefont {Frank}\ \bibnamefont
			{Saueressig}}, \bibinfo {author} {\bibfnamefont {Natalia}\ \bibnamefont
			{Alkofer}}, \bibinfo {author} {\bibfnamefont {Giulio}\ \bibnamefont
			{D'Odorico}}, \ and\ \bibinfo {author} {\bibfnamefont {Francesca}\
			\bibnamefont {Vidotto}},\ }\bibfield  {title} {\enquote {\bibinfo {title}
			{{Black holes in Asymptotically Safe Gravity}},}\ }\bibfield  {booktitle}
	{\emph {\bibinfo {booktitle} {{Proceedings, 14th International Symposium
					Frontiers of Fundamental Physics (FFP14): Marseille, France, July 15-18,
					2014}}},\ }\href {\doibase 10.22323/1.224.0174} {\bibfield  {journal}
		{\bibinfo  {journal} {PoS}\ }\textbf {\bibinfo {volume} {FFP14}},\ \bibinfo
		{pages} {174} (\bibinfo {year} {2016})},\ \Eprint
	{http://arxiv.org/abs/1503.06472} {arXiv:1503.06472 [hep-th]} \BibitemShut
	{NoStop}%
	\bibitem [{\citenamefont {Pawlowski}\ and\ \citenamefont
		{Stock}(2018)}]{Pawlowski:2018swz}%
	\BibitemOpen
	\bibfield  {author} {\bibinfo {author} {\bibfnamefont {Jan~M.}\ \bibnamefont
			{Pawlowski}}\ and\ \bibinfo {author} {\bibfnamefont {Dennis}\ \bibnamefont
			{Stock}},\ }\bibfield  {title} {\enquote {\bibinfo {title} {{Quantum-improved
					Schwarzschild-(A)dS and Kerr-(A)dS spacetimes}},}\ }\href {\doibase
		10.1103/PhysRevD.98.106008} {\bibfield  {journal} {\bibinfo  {journal} {Phys.
				Rev.}\ }\textbf {\bibinfo {volume} {D98}},\ \bibinfo {pages} {106008}
		(\bibinfo {year} {2018})},\ \Eprint {http://arxiv.org/abs/1807.10512}
	{arXiv:1807.10512 [hep-th]} \BibitemShut {NoStop}%
	\bibitem [{\citenamefont {Platania}(2019)}]{Platania:2019kyx}%
	\BibitemOpen
	\bibfield  {author} {\bibinfo {author} {\bibfnamefont {Alessia}\ \bibnamefont
			{Platania}},\ }\bibfield  {title} {\enquote {\bibinfo {title} {{Dynamical
					renormalization of black-hole spacetimes}},}\ }\href@noop {} {\  (\bibinfo
		{year} {2019})},\ \Eprint {http://arxiv.org/abs/1903.10411} {arXiv:1903.10411
		[gr-qc]} \BibitemShut {NoStop}%
	\bibitem [{\citenamefont {Donoghue}(1994)}]{Donoghue:1993eb}%
	\BibitemOpen
	\bibfield  {author} {\bibinfo {author} {\bibfnamefont {John~F.}\ \bibnamefont
			{Donoghue}},\ }\bibfield  {title} {\enquote {\bibinfo {title} {{Leading
					quantum correction to the Newtonian potential}},}\ }\href {\doibase
		10.1103/PhysRevLett.72.2996} {\bibfield  {journal} {\bibinfo  {journal}
			{Phys. Rev. Lett.}\ }\textbf {\bibinfo {volume} {72}},\ \bibinfo {pages}
		{2996--2999} (\bibinfo {year} {1994})},\ \Eprint
	{http://arxiv.org/abs/gr-qc/9310024} {arXiv:gr-qc/9310024 [gr-qc]}
	\BibitemShut {NoStop}%
	\bibitem [{\citenamefont {Donoghue}\ \emph {et~al.}(2017)\citenamefont
		{Donoghue}, \citenamefont {Ivanov},\ and\ \citenamefont
		{Shkerin}}]{Donoghue:2017pgk}%
	\BibitemOpen
	\bibfield  {author} {\bibinfo {author} {\bibfnamefont {John~F.}\ \bibnamefont
			{Donoghue}}, \bibinfo {author} {\bibfnamefont {Mikhail~M.}\ \bibnamefont
			{Ivanov}}, \ and\ \bibinfo {author} {\bibfnamefont {Andrey}\ \bibnamefont
			{Shkerin}},\ }\bibfield  {title} {\enquote {\bibinfo {title} {{EPFL Lectures
					on General Relativity as a Quantum Field Theory}},}\ }\href@noop {} {\
		(\bibinfo {year} {2017})},\ \Eprint {http://arxiv.org/abs/1702.00319}
	{arXiv:1702.00319 [hep-th]} \BibitemShut {NoStop}%
	\bibitem [{\citenamefont {Akhundov}\ and\ \citenamefont
		{Shiekh}(2008)}]{Akhundov:2006gh}%
	\BibitemOpen
	\bibfield  {author} {\bibinfo {author} {\bibfnamefont {Arif}\ \bibnamefont
			{Akhundov}}\ and\ \bibinfo {author} {\bibfnamefont {Anwar}\ \bibnamefont
			{Shiekh}},\ }\bibfield  {title} {\enquote {\bibinfo {title} {{A Review of
					Leading Quantum Gravitational Corrections to Newtonian Gravity}},}\
	}\href@noop {} {\bibfield  {journal} {\bibinfo  {journal} {Electron. J.
				Theor. Phys.}\ }\textbf {\bibinfo {volume} {5}},\ \bibinfo {pages} {1--16}
		(\bibinfo {year} {2008})},\ \Eprint {http://arxiv.org/abs/gr-qc/0611091}
	{arXiv:gr-qc/0611091 [gr-qc]} \BibitemShut {NoStop}%
	\bibitem [{\citenamefont {Bosma}\ \emph {et~al.}()\citenamefont {Bosma},
		\citenamefont {Knorr},\ and\ \citenamefont {Saueressig}}]{us2}%
	\BibitemOpen
	\bibfield  {author} {\bibinfo {author} {\bibfnamefont {Lando}\ \bibnamefont
			{Bosma}}, \bibinfo {author} {\bibfnamefont {Benjamin}\ \bibnamefont {Knorr}},
		\ and\ \bibinfo {author} {\bibfnamefont {Frank}\ \bibnamefont {Saueressig}},\
	}\href@noop {} {}\bibinfo {note} {In preparation}\BibitemShut {NoStop}%
	\bibitem [{\citenamefont {Wetterich}(1993)}]{Wetterich:1992yh}%
	\BibitemOpen
	\bibfield  {author} {\bibinfo {author} {\bibfnamefont {Christof}\
			\bibnamefont {Wetterich}},\ }\bibfield  {title} {\enquote {\bibinfo {title}
			{{Exact evolution equation for the effective potential}},}\ }\href {\doibase
		10.1016/0370-2693(93)90726-X} {\bibfield  {journal} {\bibinfo  {journal}
			{Phys.Lett.}\ }\textbf {\bibinfo {volume} {B301}},\ \bibinfo {pages} {90--94}
		(\bibinfo {year} {1993})}\BibitemShut {NoStop}%
	\bibitem [{\citenamefont {Morris}(1994)}]{Morris:1993qb}%
	\BibitemOpen
	\bibfield  {author} {\bibinfo {author} {\bibfnamefont {Tim~R.}\ \bibnamefont
			{Morris}},\ }\bibfield  {title} {\enquote {\bibinfo {title} {{The Exact
					renormalization group and approximate solutions}},}\ }\href {\doibase
		10.1142/S0217751X94000972} {\bibfield  {journal} {\bibinfo  {journal} {Int.
				J. Mod. Phys.}\ }\textbf {\bibinfo {volume} {A9}},\ \bibinfo {pages}
		{2411--2450} (\bibinfo {year} {1994})},\ \Eprint
	{http://arxiv.org/abs/hep-ph/9308265} {arXiv:hep-ph/9308265 [hep-ph]}
	\BibitemShut {NoStop}%
	\bibitem [{\citenamefont {Berges}\ \emph {et~al.}(2002)\citenamefont {Berges},
		\citenamefont {Tetradis},\ and\ \citenamefont {Wetterich}}]{Berges:2002ew}%
	\BibitemOpen
	\bibfield  {author} {\bibinfo {author} {\bibfnamefont {Juergen}\ \bibnamefont
			{Berges}}, \bibinfo {author} {\bibfnamefont {Nikolaos}\ \bibnamefont
			{Tetradis}}, \ and\ \bibinfo {author} {\bibfnamefont {Christof}\ \bibnamefont
			{Wetterich}},\ }\bibfield  {title} {\enquote {\bibinfo {title}
			{Nonperturbative renormalization flow in quantum field theory and statistical
				physics},}\ }\href {\doibase 10.1016/S0370-1573(01)00098-9} {\bibfield
		{journal} {\bibinfo  {journal} {Phys.Rept.}\ }\textbf {\bibinfo {volume}
			{363}},\ \bibinfo {pages} {223--386} (\bibinfo {year} {2002})},\ \Eprint
	{http://arxiv.org/abs/hep-ph/0005122} {arXiv:hep-ph/0005122 [hep-ph]}
	\BibitemShut {NoStop}%
	\bibitem [{\citenamefont {Berges}\ and\ \citenamefont
		{Mesterhazy}(2012)}]{Berges:2012ty}%
	\BibitemOpen
	\bibfield  {author} {\bibinfo {author} {\bibfnamefont {J.}~\bibnamefont
			{Berges}}\ and\ \bibinfo {author} {\bibfnamefont {David}\ \bibnamefont
			{Mesterhazy}},\ }\bibfield  {title} {\enquote {\bibinfo {title}
			{{Introduction to the nonequilibrium functional renormalization group}},}\
	}\href {\doibase 10.1016/j.nuclphysbps.2012.06.003} {\bibfield  {journal}
		{\bibinfo  {journal} {Nucl.Phys.Proc.Suppl.}\ }\textbf {\bibinfo {volume}
			{228}},\ \bibinfo {pages} {37--60} (\bibinfo {year} {2012})},\ \Eprint
	{http://arxiv.org/abs/1204.1489} {arXiv:1204.1489 [hep-ph]} \BibitemShut
	{NoStop}%
	\bibitem [{\citenamefont {Gies}(2012)}]{Gies:2006wv}%
	\BibitemOpen
	\bibfield  {author} {\bibinfo {author} {\bibfnamefont {Holger}\ \bibnamefont
			{Gies}},\ }\bibfield  {title} {\enquote {\bibinfo {title} {{Introduction to
					the functional RG and applications to gauge theories}},}\ }\href {\doibase
		10.1007/978-3-642-27320-9_6} {\bibfield  {journal} {\bibinfo  {journal}
			{Lect.Notes Phys.}\ }\textbf {\bibinfo {volume} {852}},\ \bibinfo {pages}
		{287--348} (\bibinfo {year} {2012})},\ \Eprint
	{http://arxiv.org/abs/hep-ph/0611146} {arXiv:hep-ph/0611146 [hep-ph]}
	\BibitemShut {NoStop}%
	\bibitem [{\citenamefont {Reuter}(1998)}]{Reuter:1996cp}%
	\BibitemOpen
	\bibfield  {author} {\bibinfo {author} {\bibfnamefont {M.}~\bibnamefont
			{Reuter}},\ }\bibfield  {title} {\enquote {\bibinfo {title} {{Nonperturbative
					evolution equation for quantum gravity}},}\ }\href {\doibase
		10.1103/PhysRevD.57.971} {\bibfield  {journal} {\bibinfo  {journal}
			{Phys.Rev.}\ }\textbf {\bibinfo {volume} {D57}},\ \bibinfo {pages} {971--985}
		(\bibinfo {year} {1998})},\ \Eprint {http://arxiv.org/abs/hep-th/9605030}
	{arXiv:hep-th/9605030 [hep-th]} \BibitemShut {NoStop}%
	\bibitem [{\citenamefont {Codello}\ \emph {et~al.}(2008)\citenamefont
		{Codello}, \citenamefont {Percacci},\ and\ \citenamefont
		{Rahmede}}]{Codello:2007bd}%
	\BibitemOpen
	\bibfield  {author} {\bibinfo {author} {\bibfnamefont {Alessandro}\
			\bibnamefont {Codello}}, \bibinfo {author} {\bibfnamefont {Roberto}\
			\bibnamefont {Percacci}}, \ and\ \bibinfo {author} {\bibfnamefont
			{Christoph}\ \bibnamefont {Rahmede}},\ }\bibfield  {title} {\enquote
		{\bibinfo {title} {{Ultraviolet properties of f(R)-gravity}},}\ }\href
	{\doibase 10.1142/S0217751X08038135} {\bibfield  {journal} {\bibinfo
			{journal} {Int.J.Mod.Phys.}\ }\textbf {\bibinfo {volume} {A23}},\ \bibinfo
		{pages} {143--150} (\bibinfo {year} {2008})},\ \Eprint
	{http://arxiv.org/abs/0705.1769} {arXiv:0705.1769 [hep-th]} \BibitemShut
	{NoStop}%
	\bibitem [{\citenamefont {Benedetti}\ \emph {et~al.}(2009)\citenamefont
		{Benedetti}, \citenamefont {Machado},\ and\ \citenamefont
		{Saueressig}}]{Benedetti:2009rx}%
	\BibitemOpen
	\bibfield  {author} {\bibinfo {author} {\bibfnamefont {Dario}\ \bibnamefont
			{Benedetti}}, \bibinfo {author} {\bibfnamefont {Pedro~F.}\ \bibnamefont
			{Machado}}, \ and\ \bibinfo {author} {\bibfnamefont {Frank}\ \bibnamefont
			{Saueressig}},\ }\bibfield  {title} {\enquote {\bibinfo {title} {{Asymptotic
					safety in higher-derivative gravity}},}\ }\href {\doibase
		10.1142/S0217732309031521} {\bibfield  {journal} {\bibinfo  {journal} {Mod.
				Phys. Lett.}\ }\textbf {\bibinfo {volume} {A24}},\ \bibinfo {pages}
		{2233--2241} (\bibinfo {year} {2009})},\ \Eprint
	{http://arxiv.org/abs/0901.2984} {arXiv:0901.2984 [hep-th]} \BibitemShut
	{NoStop}%
	\bibitem [{\citenamefont {Benedetti}\ \emph {et~al.}(2010)\citenamefont
		{Benedetti}, \citenamefont {Machado},\ and\ \citenamefont
		{Saueressig}}]{Benedetti:2009gn}%
	\BibitemOpen
	\bibfield  {author} {\bibinfo {author} {\bibfnamefont {Dario}\ \bibnamefont
			{Benedetti}}, \bibinfo {author} {\bibfnamefont {Pedro~F.}\ \bibnamefont
			{Machado}}, \ and\ \bibinfo {author} {\bibfnamefont {Frank}\ \bibnamefont
			{Saueressig}},\ }\bibfield  {title} {\enquote {\bibinfo {title} {{Taming
					perturbative divergences in asymptotically safe gravity}},}\ }\href {\doibase
		10.1016/j.nuclphysb.2009.08.023} {\bibfield  {journal} {\bibinfo  {journal}
			{Nucl. Phys.}\ }\textbf {\bibinfo {volume} {B824}},\ \bibinfo {pages}
		{168--191} (\bibinfo {year} {2010})},\ \Eprint
	{http://arxiv.org/abs/0902.4630} {arXiv:0902.4630 [hep-th]} \BibitemShut
	{NoStop}%
	\bibitem [{\citenamefont {Manrique}\ \emph {et~al.}(2011)\citenamefont
		{Manrique}, \citenamefont {Reuter},\ and\ \citenamefont
		{Saueressig}}]{Manrique:2010am}%
	\BibitemOpen
	\bibfield  {author} {\bibinfo {author} {\bibfnamefont {Elisa}\ \bibnamefont
			{Manrique}}, \bibinfo {author} {\bibfnamefont {Martin}\ \bibnamefont
			{Reuter}}, \ and\ \bibinfo {author} {\bibfnamefont {Frank}\ \bibnamefont
			{Saueressig}},\ }\bibfield  {title} {\enquote {\bibinfo {title} {{Bimetric
					Renormalization Group Flows in Quantum Einstein Gravity}},}\ }\href {\doibase
		10.1016/j.aop.2010.11.006} {\bibfield  {journal} {\bibinfo  {journal} {Annals
				Phys.}\ }\textbf {\bibinfo {volume} {326}},\ \bibinfo {pages} {463--485}
		(\bibinfo {year} {2011})},\ \Eprint {http://arxiv.org/abs/1006.0099}
	{arXiv:1006.0099 [hep-th]} \BibitemShut {NoStop}%
	\bibitem [{\citenamefont {Becker}\ and\ \citenamefont
		{Reuter}(2014)}]{Becker:2014qya}%
	\BibitemOpen
	\bibfield  {author} {\bibinfo {author} {\bibfnamefont {Daniel}\ \bibnamefont
			{Becker}}\ and\ \bibinfo {author} {\bibfnamefont {Martin}\ \bibnamefont
			{Reuter}},\ }\bibfield  {title} {\enquote {\bibinfo {title} {{En route to
					Background Independence: Broken split-symmetry, and how to restore it with
					bi-metric average actions}},}\ }\href {\doibase 10.1016/j.aop.2014.07.023}
	{\bibfield  {journal} {\bibinfo  {journal} {Annals Phys.}\ }\textbf {\bibinfo
			{volume} {350}},\ \bibinfo {pages} {225--301} (\bibinfo {year} {2014})},\
	\Eprint {http://arxiv.org/abs/1404.4537} {arXiv:1404.4537 [hep-th]}
	\BibitemShut {NoStop}%
	\bibitem [{\citenamefont {Gies}\ \emph {et~al.}(2015)\citenamefont {Gies},
		\citenamefont {Knorr},\ and\ \citenamefont {Lippoldt}}]{Gies:2015tca}%
	\BibitemOpen
	\bibfield  {author} {\bibinfo {author} {\bibfnamefont {Holger}\ \bibnamefont
			{Gies}}, \bibinfo {author} {\bibfnamefont {Benjamin}\ \bibnamefont {Knorr}},
		\ and\ \bibinfo {author} {\bibfnamefont {Stefan}\ \bibnamefont {Lippoldt}},\
	}\bibfield  {title} {\enquote {\bibinfo {title} {{Generalized Parametrization
					Dependence in Quantum Gravity}},}\ }\href {\doibase
		10.1103/PhysRevD.92.084020} {\bibfield  {journal} {\bibinfo  {journal} {Phys.
				Rev.}\ }\textbf {\bibinfo {volume} {D92}},\ \bibinfo {pages} {084020}
		(\bibinfo {year} {2015})},\ \Eprint {http://arxiv.org/abs/1507.08859}
	{arXiv:1507.08859 [hep-th]} \BibitemShut {NoStop}%
	\bibitem [{\citenamefont {Ohta}\ \emph {et~al.}(2016)\citenamefont {Ohta},
		\citenamefont {Percacci},\ and\ \citenamefont {Vacca}}]{Ohta:2015fcu}%
	\BibitemOpen
	\bibfield  {author} {\bibinfo {author} {\bibfnamefont {Nobuyoshi}\
			\bibnamefont {Ohta}}, \bibinfo {author} {\bibfnamefont {Roberto}\
			\bibnamefont {Percacci}}, \ and\ \bibinfo {author} {\bibfnamefont
			{Gian~Paolo}\ \bibnamefont {Vacca}},\ }\bibfield  {title} {\enquote {\bibinfo
			{title} {{Renormalization Group Equation and scaling solutions for f(R)
					gravity in exponential parametrization}},}\ }\href {\doibase
		10.1140/epjc/s10052-016-3895-1} {\bibfield  {journal} {\bibinfo  {journal}
			{Eur. Phys. J.}\ }\textbf {\bibinfo {volume} {C76}},\ \bibinfo {pages} {46}
		(\bibinfo {year} {2016})},\ \Eprint {http://arxiv.org/abs/1511.09393}
	{arXiv:1511.09393 [hep-th]} \BibitemShut {NoStop}%
	\bibitem [{\citenamefont {Gies}\ \emph {et~al.}(2016)\citenamefont {Gies},
		\citenamefont {Knorr}, \citenamefont {Lippoldt},\ and\ \citenamefont
		{Saueressig}}]{Gies:2016con}%
	\BibitemOpen
	\bibfield  {author} {\bibinfo {author} {\bibfnamefont {Holger}\ \bibnamefont
			{Gies}}, \bibinfo {author} {\bibfnamefont {Benjamin}\ \bibnamefont {Knorr}},
		\bibinfo {author} {\bibfnamefont {Stefan}\ \bibnamefont {Lippoldt}}, \ and\
		\bibinfo {author} {\bibfnamefont {Frank}\ \bibnamefont {Saueressig}},\
	}\bibfield  {title} {\enquote {\bibinfo {title} {{Gravitational Two-Loop
					Counterterm Is Asymptotically Safe}},}\ }\href {\doibase
		10.1103/PhysRevLett.116.211302} {\bibfield  {journal} {\bibinfo  {journal}
			{Phys. Rev. Lett.}\ }\textbf {\bibinfo {volume} {116}},\ \bibinfo {pages}
		{211302} (\bibinfo {year} {2016})},\ \Eprint
	{http://arxiv.org/abs/1601.01800} {arXiv:1601.01800 [hep-th]} \BibitemShut
	{NoStop}%
	\bibitem [{\citenamefont {Dietz}\ \emph {et~al.}(2016)\citenamefont {Dietz},
		\citenamefont {Morris},\ and\ \citenamefont {Slade}}]{Dietz:2016gzg}%
	\BibitemOpen
	\bibfield  {author} {\bibinfo {author} {\bibfnamefont {Juergen~A.}\
			\bibnamefont {Dietz}}, \bibinfo {author} {\bibfnamefont {Tim~R.}\
			\bibnamefont {Morris}}, \ and\ \bibinfo {author} {\bibfnamefont {Zoe~H.}\
			\bibnamefont {Slade}},\ }\bibfield  {title} {\enquote {\bibinfo {title}
			{{Fixed point structure of the conformal factor field in quantum gravity}},}\
	}\href {\doibase 10.1103/PhysRevD.94.124014} {\bibfield  {journal} {\bibinfo
			{journal} {Phys. Rev.}\ }\textbf {\bibinfo {volume} {D94}},\ \bibinfo {pages}
		{124014} (\bibinfo {year} {2016})},\ \Eprint
	{http://arxiv.org/abs/1605.07636} {arXiv:1605.07636 [hep-th]} \BibitemShut
	{NoStop}%
	\bibitem [{\citenamefont {Falls}\ \emph
		{et~al.}(2018{\natexlab{a}})\citenamefont {Falls}, \citenamefont {Litim},
		\citenamefont {Nikolakopoulos},\ and\ \citenamefont
		{Rahmede}}]{Falls:2016wsa}%
	\BibitemOpen
	\bibfield  {author} {\bibinfo {author} {\bibfnamefont {Kevin}\ \bibnamefont
			{Falls}}, \bibinfo {author} {\bibfnamefont {Daniel~F.}\ \bibnamefont
			{Litim}}, \bibinfo {author} {\bibfnamefont {Kostas}\ \bibnamefont
			{Nikolakopoulos}}, \ and\ \bibinfo {author} {\bibfnamefont {Christoph}\
			\bibnamefont {Rahmede}},\ }\bibfield  {title} {\enquote {\bibinfo {title}
			{{On de Sitter solutions in asymptotically safe $f(R)$ theories}},}\ }\href
	{\doibase 10.1088/1361-6382/aac440} {\bibfield  {journal} {\bibinfo
			{journal} {Class. Quant. Grav.}\ }\textbf {\bibinfo {volume} {35}},\ \bibinfo
		{pages} {135006} (\bibinfo {year} {2018}{\natexlab{a}})},\ \Eprint
	{http://arxiv.org/abs/1607.04962} {arXiv:1607.04962 [gr-qc]} \BibitemShut
	{NoStop}%
	\bibitem [{\citenamefont {Denz}\ \emph {et~al.}(2018)\citenamefont {Denz},
		\citenamefont {Pawlowski},\ and\ \citenamefont {Reichert}}]{Denz:2016qks}%
	\BibitemOpen
	\bibfield  {author} {\bibinfo {author} {\bibfnamefont {Tobias}\ \bibnamefont
			{Denz}}, \bibinfo {author} {\bibfnamefont {Jan~M.}\ \bibnamefont
			{Pawlowski}}, \ and\ \bibinfo {author} {\bibfnamefont {Manuel}\ \bibnamefont
			{Reichert}},\ }\bibfield  {title} {\enquote {\bibinfo {title} {{Towards
					apparent convergence in asymptotically safe quantum gravity}},}\ }\href
	{\doibase 10.1140/epjc/s10052-018-5806-0} {\bibfield  {journal} {\bibinfo
			{journal} {Eur. Phys. J.}\ }\textbf {\bibinfo {volume} {C78}},\ \bibinfo
		{pages} {336} (\bibinfo {year} {2018})},\ \Eprint
	{http://arxiv.org/abs/1612.07315} {arXiv:1612.07315 [hep-th]} \BibitemShut
	{NoStop}%
	\bibitem [{\citenamefont {Knorr}\ and\ \citenamefont
		{Lippoldt}(2017)}]{Knorr:2017fus}%
	\BibitemOpen
	\bibfield  {author} {\bibinfo {author} {\bibfnamefont {Benjamin}\
			\bibnamefont {Knorr}}\ and\ \bibinfo {author} {\bibfnamefont {Stefan}\
			\bibnamefont {Lippoldt}},\ }\bibfield  {title} {\enquote {\bibinfo {title}
			{{Correlation functions on a curved background}},}\ }\href {\doibase
		10.1103/PhysRevD.96.065020} {\bibfield  {journal} {\bibinfo  {journal} {Phys.
				Rev.}\ }\textbf {\bibinfo {volume} {D96}},\ \bibinfo {pages} {065020}
		(\bibinfo {year} {2017})},\ \Eprint {http://arxiv.org/abs/1707.01397}
	{arXiv:1707.01397 [hep-th]} \BibitemShut {NoStop}%
	\bibitem [{\citenamefont {Knorr}(2018)}]{Knorr:2017mhu}%
	\BibitemOpen
	\bibfield  {author} {\bibinfo {author} {\bibfnamefont {Benjamin}\
			\bibnamefont {Knorr}},\ }\bibfield  {title} {\enquote {\bibinfo {title}
			{{Infinite order quantum-gravitational correlations}},}\ }\href {\doibase
		10.1088/1361-6382/aabaa0} {\bibfield  {journal} {\bibinfo  {journal} {Class.
				Quant. Grav.}\ }\textbf {\bibinfo {volume} {35}},\ \bibinfo {pages} {115005}
		(\bibinfo {year} {2018})},\ \Eprint {http://arxiv.org/abs/1710.07055}
	{arXiv:1710.07055 [hep-th]} \BibitemShut {NoStop}%
	\bibitem [{\citenamefont {Gonzalez-Martin}\ \emph {et~al.}(2017)\citenamefont
		{Gonzalez-Martin}, \citenamefont {Morris},\ and\ \citenamefont
		{Slade}}]{Gonzalez-Martin:2017gza}%
	\BibitemOpen
	\bibfield  {author} {\bibinfo {author} {\bibfnamefont {Sergio}\ \bibnamefont
			{Gonzalez-Martin}}, \bibinfo {author} {\bibfnamefont {Tim~R.}\ \bibnamefont
			{Morris}}, \ and\ \bibinfo {author} {\bibfnamefont {Zoë~H.}\ \bibnamefont
			{Slade}},\ }\bibfield  {title} {\enquote {\bibinfo {title} {{Asymptotic
					solutions in asymptotic safety}},}\ }\href {\doibase
		10.1103/PhysRevD.95.106010} {\bibfield  {journal} {\bibinfo  {journal} {Phys.
				Rev.}\ }\textbf {\bibinfo {volume} {D95}},\ \bibinfo {pages} {106010}
		(\bibinfo {year} {2017})},\ \Eprint {http://arxiv.org/abs/1704.08873}
	{arXiv:1704.08873 [hep-th]} \BibitemShut {NoStop}%
	\bibitem [{\citenamefont {Falls}\ \emph
		{et~al.}(2018{\natexlab{b}})\citenamefont {Falls}, \citenamefont {King},
		\citenamefont {Litim}, \citenamefont {Nikolakopoulos},\ and\ \citenamefont
		{Rahmede}}]{Falls:2017lst}%
	\BibitemOpen
	\bibfield  {author} {\bibinfo {author} {\bibfnamefont {Kevin}\ \bibnamefont
			{Falls}}, \bibinfo {author} {\bibfnamefont {Callum~R.}\ \bibnamefont {King}},
		\bibinfo {author} {\bibfnamefont {Daniel~F.}\ \bibnamefont {Litim}}, \bibinfo
		{author} {\bibfnamefont {Kostas}\ \bibnamefont {Nikolakopoulos}}, \ and\
		\bibinfo {author} {\bibfnamefont {Christoph}\ \bibnamefont {Rahmede}},\
	}\bibfield  {title} {\enquote {\bibinfo {title} {{Asymptotic safety of
					quantum gravity beyond Ricci scalars}},}\ }\href {\doibase
		10.1103/PhysRevD.97.086006} {\bibfield  {journal} {\bibinfo  {journal} {Phys.
				Rev.}\ }\textbf {\bibinfo {volume} {D97}},\ \bibinfo {pages} {086006}
		(\bibinfo {year} {2018}{\natexlab{b}})},\ \Eprint
	{http://arxiv.org/abs/1801.00162} {arXiv:1801.00162 [hep-th]} \BibitemShut
	{NoStop}%
	\bibitem [{\citenamefont {Christiansen}\ \emph
		{et~al.}(2018{\natexlab{a}})\citenamefont {Christiansen}, \citenamefont
		{Falls}, \citenamefont {Pawlowski},\ and\ \citenamefont
		{Reichert}}]{Christiansen:2017bsy}%
	\BibitemOpen
	\bibfield  {author} {\bibinfo {author} {\bibfnamefont {Nicolai}\ \bibnamefont
			{Christiansen}}, \bibinfo {author} {\bibfnamefont {Kevin}\ \bibnamefont
			{Falls}}, \bibinfo {author} {\bibfnamefont {Jan~M.}\ \bibnamefont
			{Pawlowski}}, \ and\ \bibinfo {author} {\bibfnamefont {Manuel}\ \bibnamefont
			{Reichert}},\ }\bibfield  {title} {\enquote {\bibinfo {title} {{Curvature
					dependence of quantum gravity}},}\ }\href {\doibase
		10.1103/PhysRevD.97.046007} {\bibfield  {journal} {\bibinfo  {journal} {Phys.
				Rev.}\ }\textbf {\bibinfo {volume} {D97}},\ \bibinfo {pages} {046007}
		(\bibinfo {year} {2018}{\natexlab{a}})},\ \Eprint
	{http://arxiv.org/abs/1711.09259} {arXiv:1711.09259 [hep-th]} \BibitemShut
	{NoStop}%
	\bibitem [{\citenamefont {Alkofer}\ and\ \citenamefont
		{Saueressig}(2018)}]{Alkofer:2018fxj}%
	\BibitemOpen
	\bibfield  {author} {\bibinfo {author} {\bibfnamefont {Natália}\
			\bibnamefont {Alkofer}}\ and\ \bibinfo {author} {\bibfnamefont {Frank}\
			\bibnamefont {Saueressig}},\ }\bibfield  {title} {\enquote {\bibinfo {title}
			{{Asymptotically safe $f(R)$-gravity coupled to matter I: the polynomial
					case}},}\ }\href {\doibase 10.1016/j.aop.2018.07.017} {\bibfield  {journal}
		{\bibinfo  {journal} {Annals Phys.}\ }\textbf {\bibinfo {volume} {396}},\
		\bibinfo {pages} {173--201} (\bibinfo {year} {2018})},\ \Eprint
	{http://arxiv.org/abs/1802.00498} {arXiv:1802.00498 [hep-th]} \BibitemShut
	{NoStop}%
	\bibitem [{\citenamefont {Alkofer}(2019)}]{Alkofer:2018baq}%
	\BibitemOpen
	\bibfield  {author} {\bibinfo {author} {\bibfnamefont {Natália}\
			\bibnamefont {Alkofer}},\ }\bibfield  {title} {\enquote {\bibinfo {title}
			{{Asymptotically safe $f(R)$-gravity coupled to matter II: Global
					solutions}},}\ }\href {\doibase 10.1016/j.physletb.2018.12.061} {\bibfield
		{journal} {\bibinfo  {journal} {Phys. Lett.}\ }\textbf {\bibinfo {volume}
			{B789}},\ \bibinfo {pages} {480--487} (\bibinfo {year} {2019})},\ \Eprint
	{http://arxiv.org/abs/1809.06162} {arXiv:1809.06162 [hep-th]} \BibitemShut
	{NoStop}%
	\bibitem [{\citenamefont {Gubitosi}\ \emph {et~al.}(2018)\citenamefont
		{Gubitosi}, \citenamefont {Ooijer}, \citenamefont {Ripken},\ and\
		\citenamefont {Saueressig}}]{Gubitosi:2018gsl}%
	\BibitemOpen
	\bibfield  {author} {\bibinfo {author} {\bibfnamefont {Giulia}\ \bibnamefont
			{Gubitosi}}, \bibinfo {author} {\bibfnamefont {Robin}\ \bibnamefont
			{Ooijer}}, \bibinfo {author} {\bibfnamefont {Chris}\ \bibnamefont {Ripken}},
		\ and\ \bibinfo {author} {\bibfnamefont {Frank}\ \bibnamefont {Saueressig}},\
	}\bibfield  {title} {\enquote {\bibinfo {title} {{Consistent early and late
					time cosmology from the RG flow of gravity}},}\ }\href {\doibase
		10.1088/1475-7516/2018/12/004} {\bibfield  {journal} {\bibinfo  {journal}
			{JCAP}\ }\textbf {\bibinfo {volume} {1812}},\ \bibinfo {pages} {004}
		(\bibinfo {year} {2018})},\ \Eprint {http://arxiv.org/abs/1806.10147}
	{arXiv:1806.10147 [hep-th]} \BibitemShut {NoStop}%
	\bibitem [{\citenamefont {Falls}\ \emph
		{et~al.}(2018{\natexlab{c}})\citenamefont {Falls}, \citenamefont {Litim},\
		and\ \citenamefont {Schröder}}]{Falls:2018ylp}%
	\BibitemOpen
	\bibfield  {author} {\bibinfo {author} {\bibfnamefont {Kevin~G.}\
			\bibnamefont {Falls}}, \bibinfo {author} {\bibfnamefont {Daniel~F.}\
			\bibnamefont {Litim}}, \ and\ \bibinfo {author} {\bibfnamefont {Jan}\
			\bibnamefont {Schröder}},\ }\bibfield  {title} {\enquote {\bibinfo {title}
			{{Aspects of asymptotic safety for quantum gravity}},}\ }\href@noop {} {\
		(\bibinfo {year} {2018}{\natexlab{c}})},\ \Eprint
	{http://arxiv.org/abs/1810.08550} {arXiv:1810.08550 [gr-qc]} \BibitemShut
	{NoStop}%
	\bibitem [{\citenamefont {Eichhorn}\ \emph
		{et~al.}(2018{\natexlab{a}})\citenamefont {Eichhorn}, \citenamefont
		{Lippoldt}, \citenamefont {Pawlowski}, \citenamefont {Reichert},\ and\
		\citenamefont {Schiffer}}]{Eichhorn:2018ydy}%
	\BibitemOpen
	\bibfield  {author} {\bibinfo {author} {\bibfnamefont {Astrid}\ \bibnamefont
			{Eichhorn}}, \bibinfo {author} {\bibfnamefont {Stefan}\ \bibnamefont
			{Lippoldt}}, \bibinfo {author} {\bibfnamefont {Jan~M.}\ \bibnamefont
			{Pawlowski}}, \bibinfo {author} {\bibfnamefont {Manuel}\ \bibnamefont
			{Reichert}}, \ and\ \bibinfo {author} {\bibfnamefont {Marc}\ \bibnamefont
			{Schiffer}},\ }\bibfield  {title} {\enquote {\bibinfo {title} {{How
					perturbative is quantum gravity?}}}\ }\href {\doibase
		10.1016/j.physletb.2019.01.071} {\  (\bibinfo {year} {2018}{\natexlab{a}}),\
		10.1016/j.physletb.2019.01.071},\ \Eprint {http://arxiv.org/abs/1810.02828}
	{arXiv:1810.02828 [hep-th]} \BibitemShut {NoStop}%
	\bibitem [{\citenamefont {Eichhorn}\ \emph
		{et~al.}(2018{\natexlab{b}})\citenamefont {Eichhorn}, \citenamefont {Labus},
		\citenamefont {Pawlowski},\ and\ \citenamefont
		{Reichert}}]{Eichhorn:2018akn}%
	\BibitemOpen
	\bibfield  {author} {\bibinfo {author} {\bibfnamefont {Astrid}\ \bibnamefont
			{Eichhorn}}, \bibinfo {author} {\bibfnamefont {Peter}\ \bibnamefont {Labus}},
		\bibinfo {author} {\bibfnamefont {Jan~M.}\ \bibnamefont {Pawlowski}}, \ and\
		\bibinfo {author} {\bibfnamefont {Manuel}\ \bibnamefont {Reichert}},\
	}\bibfield  {title} {\enquote {\bibinfo {title} {{Effective universality in
					quantum gravity}},}\ }\href {\doibase 10.21468/SciPostPhys.5.4.031}
	{\bibfield  {journal} {\bibinfo  {journal} {SciPost Phys.}\ }\textbf
		{\bibinfo {volume} {5}},\ \bibinfo {pages} {031} (\bibinfo {year}
		{2018}{\natexlab{b}})},\ \Eprint {http://arxiv.org/abs/1804.00012}
	{arXiv:1804.00012 [hep-th]} \BibitemShut {NoStop}%
	\bibitem [{\citenamefont {Don\`a}\ \emph {et~al.}(2014)\citenamefont {Don\`a},
		\citenamefont {Eichhorn},\ and\ \citenamefont {Percacci}}]{Dona:2013qba}%
	\BibitemOpen
	\bibfield  {author} {\bibinfo {author} {\bibfnamefont {Pietro}\ \bibnamefont
			{Don\`a}}, \bibinfo {author} {\bibfnamefont {Astrid}\ \bibnamefont
			{Eichhorn}}, \ and\ \bibinfo {author} {\bibfnamefont {Roberto}\ \bibnamefont
			{Percacci}},\ }\bibfield  {title} {\enquote {\bibinfo {title} {{Matter
					matters in asymptotically safe quantum gravity}},}\ }\href {\doibase
		10.1103/PhysRevD.89.084035} {\bibfield  {journal} {\bibinfo  {journal}
			{Phys.Rev.}\ }\textbf {\bibinfo {volume} {D89}},\ \bibinfo {pages} {084035}
		(\bibinfo {year} {2014})},\ \Eprint {http://arxiv.org/abs/1311.2898}
	{arXiv:1311.2898 [hep-th]} \BibitemShut {NoStop}%
	\bibitem [{\citenamefont {Meibohm}\ and\ \citenamefont
		{Pawlowski}(2016)}]{Meibohm:2016mkp}%
	\BibitemOpen
	\bibfield  {author} {\bibinfo {author} {\bibfnamefont {Jan}\ \bibnamefont
			{Meibohm}}\ and\ \bibinfo {author} {\bibfnamefont {Jan~M.}\ \bibnamefont
			{Pawlowski}},\ }\bibfield  {title} {\enquote {\bibinfo {title} {{Chiral
					fermions in asymptotically safe quantum gravity}},}\ }\href {\doibase
		10.1140/epjc/s10052-016-4132-7} {\bibfield  {journal} {\bibinfo  {journal}
			{Eur. Phys. J.}\ }\textbf {\bibinfo {volume} {C76}},\ \bibinfo {pages} {285}
		(\bibinfo {year} {2016})},\ \Eprint {http://arxiv.org/abs/1601.04597}
	{arXiv:1601.04597 [hep-th]} \BibitemShut {NoStop}%
	\bibitem [{\citenamefont {Eichhorn}\ \emph {et~al.}(2016)\citenamefont
		{Eichhorn}, \citenamefont {Held},\ and\ \citenamefont
		{Pawlowski}}]{Eichhorn:2016esv}%
	\BibitemOpen
	\bibfield  {author} {\bibinfo {author} {\bibfnamefont {Astrid}\ \bibnamefont
			{Eichhorn}}, \bibinfo {author} {\bibfnamefont {Aaron}\ \bibnamefont {Held}},
		\ and\ \bibinfo {author} {\bibfnamefont {Jan~M.}\ \bibnamefont {Pawlowski}},\
	}\bibfield  {title} {\enquote {\bibinfo {title} {{Quantum-gravity effects on
					a Higgs-Yukawa model}},}\ }\href {\doibase 10.1103/PhysRevD.94.104027}
	{\bibfield  {journal} {\bibinfo  {journal} {Phys. Rev.}\ }\textbf {\bibinfo
			{volume} {D94}},\ \bibinfo {pages} {104027} (\bibinfo {year} {2016})},\
	\Eprint {http://arxiv.org/abs/1604.02041} {arXiv:1604.02041 [hep-th]}
	\BibitemShut {NoStop}%
	\bibitem [{\citenamefont {Eichhorn}\ and\ \citenamefont
		{Lippoldt}(2017)}]{Eichhorn:2016vvy}%
	\BibitemOpen
	\bibfield  {author} {\bibinfo {author} {\bibfnamefont {Astrid}\ \bibnamefont
			{Eichhorn}}\ and\ \bibinfo {author} {\bibfnamefont {Stefan}\ \bibnamefont
			{Lippoldt}},\ }\bibfield  {title} {\enquote {\bibinfo {title} {{Quantum
					gravity and Standard-Model-like fermions}},}\ }\href {\doibase
		10.1016/j.physletb.2017.01.064} {\bibfield  {journal} {\bibinfo  {journal}
			{Phys. Lett.}\ }\textbf {\bibinfo {volume} {B767}},\ \bibinfo {pages}
		{142--146} (\bibinfo {year} {2017})},\ \Eprint
	{http://arxiv.org/abs/1611.05878} {arXiv:1611.05878 [gr-qc]} \BibitemShut
	{NoStop}%
	\bibitem [{\citenamefont {Christiansen}\ \emph
		{et~al.}(2018{\natexlab{b}})\citenamefont {Christiansen}, \citenamefont
		{Litim}, \citenamefont {Pawlowski},\ and\ \citenamefont
		{Reichert}}]{Christiansen:2017cxa}%
	\BibitemOpen
	\bibfield  {author} {\bibinfo {author} {\bibfnamefont {Nicolai}\ \bibnamefont
			{Christiansen}}, \bibinfo {author} {\bibfnamefont {Daniel~F.}\ \bibnamefont
			{Litim}}, \bibinfo {author} {\bibfnamefont {Jan~M.}\ \bibnamefont
			{Pawlowski}}, \ and\ \bibinfo {author} {\bibfnamefont {Manuel}\ \bibnamefont
			{Reichert}},\ }\bibfield  {title} {\enquote {\bibinfo {title} {{Asymptotic
					safety of gravity with matter}},}\ }\href {\doibase
		10.1103/PhysRevD.97.106012} {\bibfield  {journal} {\bibinfo  {journal} {Phys.
				Rev.}\ }\textbf {\bibinfo {volume} {D97}},\ \bibinfo {pages} {106012}
		(\bibinfo {year} {2018}{\natexlab{b}})},\ \Eprint
	{http://arxiv.org/abs/1710.04669} {arXiv:1710.04669 [hep-th]} \BibitemShut
	{NoStop}%
	\bibitem [{\citenamefont {Christiansen}\ and\ \citenamefont
		{Eichhorn}(2017)}]{Christiansen:2017gtg}%
	\BibitemOpen
	\bibfield  {author} {\bibinfo {author} {\bibfnamefont {Nicolai}\ \bibnamefont
			{Christiansen}}\ and\ \bibinfo {author} {\bibfnamefont {Astrid}\ \bibnamefont
			{Eichhorn}},\ }\bibfield  {title} {\enquote {\bibinfo {title} {{An
					asymptotically safe solution to the U(1) triviality problem}},}\ }\href
	{\doibase 10.1016/j.physletb.2017.04.047} {\bibfield  {journal} {\bibinfo
			{journal} {Phys. Lett.}\ }\textbf {\bibinfo {volume} {B770}},\ \bibinfo
		{pages} {154--160} (\bibinfo {year} {2017})},\ \Eprint
	{http://arxiv.org/abs/1702.07724} {arXiv:1702.07724 [hep-th]} \BibitemShut
	{NoStop}%
	\bibitem [{\citenamefont {Eichhorn}\ and\ \citenamefont
		{Held}(2018{\natexlab{a}})}]{Eichhorn:2017ylw}%
	\BibitemOpen
	\bibfield  {author} {\bibinfo {author} {\bibfnamefont {Astrid}\ \bibnamefont
			{Eichhorn}}\ and\ \bibinfo {author} {\bibfnamefont {Aaron}\ \bibnamefont
			{Held}},\ }\bibfield  {title} {\enquote {\bibinfo {title} {{Top mass from
					asymptotic safety}},}\ }\href {\doibase 10.1016/j.physletb.2017.12.040}
	{\bibfield  {journal} {\bibinfo  {journal} {Phys. Lett.}\ }\textbf {\bibinfo
			{volume} {B777}},\ \bibinfo {pages} {217--221} (\bibinfo {year}
		{2018}{\natexlab{a}})},\ \Eprint {http://arxiv.org/abs/1707.01107}
	{arXiv:1707.01107 [hep-th]} \BibitemShut {NoStop}%
	\bibitem [{\citenamefont {Eichhorn}\ and\ \citenamefont
		{Versteegen}(2018)}]{Eichhorn:2017lry}%
	\BibitemOpen
	\bibfield  {author} {\bibinfo {author} {\bibfnamefont {Astrid}\ \bibnamefont
			{Eichhorn}}\ and\ \bibinfo {author} {\bibfnamefont {Fleur}\ \bibnamefont
			{Versteegen}},\ }\bibfield  {title} {\enquote {\bibinfo {title} {{Upper bound
					on the Abelian gauge coupling from asymptotic safety}},}\ }\href {\doibase
		10.1007/JHEP01(2018)030} {\bibfield  {journal} {\bibinfo  {journal} {JHEP}\
		}\textbf {\bibinfo {volume} {01}},\ \bibinfo {pages} {030} (\bibinfo {year}
		{2018})},\ \Eprint {http://arxiv.org/abs/1709.07252} {arXiv:1709.07252
		[hep-th]} \BibitemShut {NoStop}%
	\bibitem [{\citenamefont {Eichhorn}\ \emph
		{et~al.}(2018{\natexlab{c}})\citenamefont {Eichhorn}, \citenamefont {Held},\
		and\ \citenamefont {Wetterich}}]{Eichhorn:2017muy}%
	\BibitemOpen
	\bibfield  {author} {\bibinfo {author} {\bibfnamefont {Astrid}\ \bibnamefont
			{Eichhorn}}, \bibinfo {author} {\bibfnamefont {Aaron}\ \bibnamefont {Held}},
		\ and\ \bibinfo {author} {\bibfnamefont {Christof}\ \bibnamefont
			{Wetterich}},\ }\bibfield  {title} {\enquote {\bibinfo {title}
			{{Quantum-gravity predictions for the fine-structure constant}},}\ }\href
	{\doibase 10.1016/j.physletb.2018.05.016} {\bibfield  {journal} {\bibinfo
			{journal} {Phys. Lett.}\ }\textbf {\bibinfo {volume} {B782}},\ \bibinfo
		{pages} {198--201} (\bibinfo {year} {2018}{\natexlab{c}})},\ \Eprint
	{http://arxiv.org/abs/1711.02949} {arXiv:1711.02949 [hep-th]} \BibitemShut
	{NoStop}%
	\bibitem [{\citenamefont {Eichhorn}\ \emph {et~al.}(2019)\citenamefont
		{Eichhorn}, \citenamefont {Lippoldt},\ and\ \citenamefont
		{Schiffer}}]{Eichhorn:2018nda}%
	\BibitemOpen
	\bibfield  {author} {\bibinfo {author} {\bibfnamefont {Astrid}\ \bibnamefont
			{Eichhorn}}, \bibinfo {author} {\bibfnamefont {Stefan}\ \bibnamefont
			{Lippoldt}}, \ and\ \bibinfo {author} {\bibfnamefont {Marc}\ \bibnamefont
			{Schiffer}},\ }\bibfield  {title} {\enquote {\bibinfo {title} {{Zooming in on
					fermions and quantum gravity}},}\ }\href {\doibase
		10.1103/PhysRevD.99.086002} {\bibfield  {journal} {\bibinfo  {journal} {Phys.
				Rev.}\ }\textbf {\bibinfo {volume} {D99}},\ \bibinfo {pages} {086002}
		(\bibinfo {year} {2019})},\ \Eprint {http://arxiv.org/abs/1812.08782}
	{arXiv:1812.08782 [hep-th]} \BibitemShut {NoStop}%
	\bibitem [{\citenamefont {Pawlowski}\ \emph {et~al.}(2018)\citenamefont
		{Pawlowski}, \citenamefont {Reichert}, \citenamefont {Wetterich},\ and\
		\citenamefont {Yamada}}]{Pawlowski:2018ixd}%
	\BibitemOpen
	\bibfield  {author} {\bibinfo {author} {\bibfnamefont {Jan~M.}\ \bibnamefont
			{Pawlowski}}, \bibinfo {author} {\bibfnamefont {Manuel}\ \bibnamefont
			{Reichert}}, \bibinfo {author} {\bibfnamefont {Christof}\ \bibnamefont
			{Wetterich}}, \ and\ \bibinfo {author} {\bibfnamefont {Masatoshi}\
			\bibnamefont {Yamada}},\ }\bibfield  {title} {\enquote {\bibinfo {title}
			{{Higgs scalar potential in asymptotically safe quantum gravity}},}\
	}\href@noop {} {\  (\bibinfo {year} {2018})},\ \Eprint
	{http://arxiv.org/abs/1811.11706} {arXiv:1811.11706 [hep-th]} \BibitemShut
	{NoStop}%
	\bibitem [{\citenamefont {Eichhorn}\ and\ \citenamefont
		{Held}(2018{\natexlab{b}})}]{Eichhorn:2018whv}%
	\BibitemOpen
	\bibfield  {author} {\bibinfo {author} {\bibfnamefont {Astrid}\ \bibnamefont
			{Eichhorn}}\ and\ \bibinfo {author} {\bibfnamefont {Aaron}\ \bibnamefont
			{Held}},\ }\bibfield  {title} {\enquote {\bibinfo {title} {{Mass difference
					for charged quarks from asymptotically safe quantum gravity}},}\ }\href
	{\doibase 10.1103/PhysRevLett.121.151302} {\bibfield  {journal} {\bibinfo
			{journal} {Phys. Rev. Lett.}\ }\textbf {\bibinfo {volume} {121}},\ \bibinfo
		{pages} {151302} (\bibinfo {year} {2018}{\natexlab{b}})},\ \Eprint
	{http://arxiv.org/abs/1803.04027} {arXiv:1803.04027 [hep-th]} \BibitemShut
	{NoStop}%
	\bibitem [{\citenamefont {Christiansen}\ \emph {et~al.}(2016)\citenamefont
		{Christiansen}, \citenamefont {Knorr}, \citenamefont {Pawlowski},\ and\
		\citenamefont {Rodigast}}]{Christiansen:2014raa}%
	\BibitemOpen
	\bibfield  {author} {\bibinfo {author} {\bibfnamefont {Nicolai}\ \bibnamefont
			{Christiansen}}, \bibinfo {author} {\bibfnamefont {Benjamin}\ \bibnamefont
			{Knorr}}, \bibinfo {author} {\bibfnamefont {Jan~M.}\ \bibnamefont
			{Pawlowski}}, \ and\ \bibinfo {author} {\bibfnamefont {Andreas}\ \bibnamefont
			{Rodigast}},\ }\bibfield  {title} {\enquote {\bibinfo {title} {{Global Flows
					in Quantum Gravity}},}\ }\href {\doibase 10.1103/PhysRevD.93.044036}
	{\bibfield  {journal} {\bibinfo  {journal} {Phys. Rev.}\ }\textbf {\bibinfo
			{volume} {D93}},\ \bibinfo {pages} {044036} (\bibinfo {year} {2016})},\
	\Eprint {http://arxiv.org/abs/1403.1232} {arXiv:1403.1232 [hep-th]}
	\BibitemShut {NoStop}%
	\bibitem [{\citenamefont {Christiansen}\ \emph {et~al.}(2015)\citenamefont
		{Christiansen}, \citenamefont {Knorr}, \citenamefont {Meibohm}, \citenamefont
		{Pawlowski},\ and\ \citenamefont {Reichert}}]{Christiansen:2015rva}%
	\BibitemOpen
	\bibfield  {author} {\bibinfo {author} {\bibfnamefont {Nicolai}\ \bibnamefont
			{Christiansen}}, \bibinfo {author} {\bibfnamefont {Benjamin}\ \bibnamefont
			{Knorr}}, \bibinfo {author} {\bibfnamefont {Jan}\ \bibnamefont {Meibohm}},
		\bibinfo {author} {\bibfnamefont {Jan~M.}\ \bibnamefont {Pawlowski}}, \ and\
		\bibinfo {author} {\bibfnamefont {Manuel}\ \bibnamefont {Reichert}},\
	}\bibfield  {title} {\enquote {\bibinfo {title} {{Local Quantum Gravity}},}\
	}\href {\doibase 10.1103/PhysRevD.92.121501} {\bibfield  {journal} {\bibinfo
			{journal} {Phys. Rev.}\ }\textbf {\bibinfo {volume} {D92}},\ \bibinfo {pages}
		{121501} (\bibinfo {year} {2015})},\ \Eprint
	{http://arxiv.org/abs/1506.07016} {arXiv:1506.07016 [hep-th]} \BibitemShut
	{NoStop}%
	\bibitem [{\citenamefont {Weinberg}(1976)}]{Weinberg:1976xy}%
	\BibitemOpen
	\bibfield  {author} {\bibinfo {author} {\bibfnamefont {Steven}\ \bibnamefont
			{Weinberg}},\ }\bibfield  {title} {\enquote {\bibinfo {title} {{Critical
					Phenomena for Field Theorists}},}\ }in\ \href@noop {} {\emph {\bibinfo
			{booktitle} {{Erice Subnucl.Phys.1976:1}}}}\ (\bibinfo {year} {1976})\
	p.~\bibinfo {pages} {1}\BibitemShut {NoStop}%
	\bibitem [{\citenamefont {Reuter}\ and\ \citenamefont
		{Weyer}(2009{\natexlab{a}})}]{Reuter:2008wj}%
	\BibitemOpen
	\bibfield  {author} {\bibinfo {author} {\bibfnamefont {Martin}\ \bibnamefont
			{Reuter}}\ and\ \bibinfo {author} {\bibfnamefont {Holger}\ \bibnamefont
			{Weyer}},\ }\bibfield  {title} {\enquote {\bibinfo {title} {{Background
					Independence and Asymptotic Safety in Conformally Reduced Gravity}},}\ }\href
	{\doibase 10.1103/PhysRevD.79.105005} {\bibfield  {journal} {\bibinfo
			{journal} {Phys. Rev.}\ }\textbf {\bibinfo {volume} {D79}},\ \bibinfo {pages}
		{105005} (\bibinfo {year} {2009}{\natexlab{a}})},\ \Eprint
	{http://arxiv.org/abs/0801.3287} {arXiv:0801.3287 [hep-th]} \BibitemShut
	{NoStop}%
	\bibitem [{\citenamefont {Reuter}\ and\ \citenamefont
		{Weyer}(2009{\natexlab{b}})}]{Reuter:2008qx}%
	\BibitemOpen
	\bibfield  {author} {\bibinfo {author} {\bibfnamefont {Martin}\ \bibnamefont
			{Reuter}}\ and\ \bibinfo {author} {\bibfnamefont {Holger}\ \bibnamefont
			{Weyer}},\ }\bibfield  {title} {\enquote {\bibinfo {title} {{Conformal sector
					of Quantum Einstein Gravity in the local potential approximation:
					Non-Gaussian fixed point and a phase of unbroken diffeomorphism
					invariance}},}\ }\href {\doibase 10.1103/PhysRevD.80.025001} {\bibfield
		{journal} {\bibinfo  {journal} {Phys. Rev.}\ }\textbf {\bibinfo {volume}
			{D80}},\ \bibinfo {pages} {025001} (\bibinfo {year} {2009}{\natexlab{b}})},\
	\Eprint {http://arxiv.org/abs/0804.1475} {arXiv:0804.1475 [hep-th]}
	\BibitemShut {NoStop}%
	\bibitem [{\citenamefont {Demmel}\ \emph {et~al.}(2012)\citenamefont {Demmel},
		\citenamefont {Saueressig},\ and\ \citenamefont {Zanusso}}]{Demmel:2012ub}%
	\BibitemOpen
	\bibfield  {author} {\bibinfo {author} {\bibfnamefont {Maximilian}\
			\bibnamefont {Demmel}}, \bibinfo {author} {\bibfnamefont {Frank}\
			\bibnamefont {Saueressig}}, \ and\ \bibinfo {author} {\bibfnamefont {Omar}\
			\bibnamefont {Zanusso}},\ }\bibfield  {title} {\enquote {\bibinfo {title}
			{{Fixed-Functionals of three-dimensional Quantum Einstein Gravity}},}\ }\href
	{\doibase 10.1007/JHEP11(2012)131} {\bibfield  {journal} {\bibinfo  {journal}
			{JHEP}\ }\textbf {\bibinfo {volume} {11}},\ \bibinfo {pages} {131} (\bibinfo
		{year} {2012})},\ \Eprint {http://arxiv.org/abs/1208.2038} {arXiv:1208.2038
		[hep-th]} \BibitemShut {NoStop}%
	\bibitem [{\citenamefont {Demmel}\ \emph {et~al.}(2014)\citenamefont {Demmel},
		\citenamefont {Saueressig},\ and\ \citenamefont {Zanusso}}]{Demmel:2014sga}%
	\BibitemOpen
	\bibfield  {author} {\bibinfo {author} {\bibfnamefont {Maximilian}\
			\bibnamefont {Demmel}}, \bibinfo {author} {\bibfnamefont {Frank}\
			\bibnamefont {Saueressig}}, \ and\ \bibinfo {author} {\bibfnamefont {Omar}\
			\bibnamefont {Zanusso}},\ }\bibfield  {title} {\enquote {\bibinfo {title}
			{{RG flows of Quantum Einstein Gravity on maximally symmetric spaces}},}\
	}\href {\doibase 10.1007/JHEP06(2014)026} {\bibfield  {journal} {\bibinfo
			{journal} {JHEP}\ }\textbf {\bibinfo {volume} {06}},\ \bibinfo {pages} {026}
		(\bibinfo {year} {2014})},\ \Eprint {http://arxiv.org/abs/1401.5495}
	{arXiv:1401.5495 [hep-th]} \BibitemShut {NoStop}%
	\bibitem [{\citenamefont {Demmel}\ \emph {et~al.}(2015)\citenamefont {Demmel},
		\citenamefont {Saueressig},\ and\ \citenamefont {Zanusso}}]{Demmel:2015oqa}%
	\BibitemOpen
	\bibfield  {author} {\bibinfo {author} {\bibfnamefont {Maximilian}\
			\bibnamefont {Demmel}}, \bibinfo {author} {\bibfnamefont {Frank}\
			\bibnamefont {Saueressig}}, \ and\ \bibinfo {author} {\bibfnamefont {Omar}\
			\bibnamefont {Zanusso}},\ }\bibfield  {title} {\enquote {\bibinfo {title} {{A
					proper fixed functional for four-dimensional Quantum Einstein Gravity}},}\
	}\href {\doibase 10.1007/JHEP08(2015)113} {\bibfield  {journal} {\bibinfo
			{journal} {JHEP}\ }\textbf {\bibinfo {volume} {08}},\ \bibinfo {pages} {113}
		(\bibinfo {year} {2015})},\ \Eprint {http://arxiv.org/abs/1504.07656}
	{arXiv:1504.07656 [hep-th]} \BibitemShut {NoStop}%
	\bibitem [{\citenamefont {Dietz}\ and\ \citenamefont
		{Morris}(2015)}]{Dietz:2015owa}%
	\BibitemOpen
	\bibfield  {author} {\bibinfo {author} {\bibfnamefont {Juergen~A.}\
			\bibnamefont {Dietz}}\ and\ \bibinfo {author} {\bibfnamefont {Tim~R.}\
			\bibnamefont {Morris}},\ }\bibfield  {title} {\enquote {\bibinfo {title}
			{{Background independent exact renormalization group for conformally reduced
					gravity}},}\ }\href {\doibase 10.1007/JHEP04(2015)118} {\bibfield  {journal}
		{\bibinfo  {journal} {JHEP}\ }\textbf {\bibinfo {volume} {04}},\ \bibinfo
		{pages} {118} (\bibinfo {year} {2015})},\ \Eprint
	{http://arxiv.org/abs/1502.07396} {arXiv:1502.07396 [hep-th]} \BibitemShut
	{NoStop}%
	\bibitem [{\citenamefont {Codello}\ \emph {et~al.}(2009)\citenamefont
		{Codello}, \citenamefont {Percacci},\ and\ \citenamefont
		{Rahmede}}]{Codello:2008vh}%
	\BibitemOpen
	\bibfield  {author} {\bibinfo {author} {\bibfnamefont {Alessandro}\
			\bibnamefont {Codello}}, \bibinfo {author} {\bibfnamefont {Roberto}\
			\bibnamefont {Percacci}}, \ and\ \bibinfo {author} {\bibfnamefont
			{Christoph}\ \bibnamefont {Rahmede}},\ }\bibfield  {title} {\enquote
		{\bibinfo {title} {{Investigating the Ultraviolet Properties of Gravity with
					a Wilsonian Renormalization Group Equation}},}\ }\href {\doibase
		10.1016/j.aop.2008.08.008} {\bibfield  {journal} {\bibinfo  {journal} {Annals
				Phys.}\ }\textbf {\bibinfo {volume} {324}},\ \bibinfo {pages} {414--469}
		(\bibinfo {year} {2009})},\ \Eprint {http://arxiv.org/abs/0805.2909}
	{arXiv:0805.2909 [hep-th]} \BibitemShut {NoStop}%
	\bibitem [{\citenamefont {Codello}\ and\ \citenamefont
		{Zanusso}(2013)}]{Codello:2012kq}%
	\BibitemOpen
	\bibfield  {author} {\bibinfo {author} {\bibfnamefont {Alessandro}\
			\bibnamefont {Codello}}\ and\ \bibinfo {author} {\bibfnamefont {Omar}\
			\bibnamefont {Zanusso}},\ }\bibfield  {title} {\enquote {\bibinfo {title}
			{{On the non-local heat kernel expansion}},}\ }\href {\doibase
		10.1063/1.4776234} {\bibfield  {journal} {\bibinfo  {journal} {J. Math.
				Phys.}\ }\textbf {\bibinfo {volume} {54}},\ \bibinfo {pages} {013513}
		(\bibinfo {year} {2013})},\ \Eprint {http://arxiv.org/abs/1203.2034}
	{arXiv:1203.2034 [math-ph]} \BibitemShut {NoStop}%
	\bibitem{Meibohm:2015twa} 
	J.~Meibohm, J.~M.~Pawlowski and M.~Reichert,
	``Asymptotic safety of gravity-matter systems,
	Phys.\ Rev.\ D {\bf 93}, 084035 (2016),
	arXiv:1510.07018 [hep-th].
	\bibitem [{\citenamefont {Knorr}\ \emph {et~al.}()\citenamefont {Knorr},
		\citenamefont {Ripken},\ and\ \citenamefont {Saueressig}}]{Knorr:CQG}%
	\BibitemOpen
	\bibfield  {author} {\bibinfo {author} {\bibfnamefont {Benjamin}\
			\bibnamefont {Knorr}}, \bibinfo {author} {\bibfnamefont {Chris}\ \bibnamefont
			{Ripken}}, \ and\ \bibinfo {author} {\bibfnamefont {Frank}\ \bibnamefont
			{Saueressig}},\ }\href@noop {} {}\bibfield  {title} {\enquote {\bibinfo {title}
			{{Form Factors in Asymptotic Safety: conceptual ideas and computational toolbox}},}\ } \Eprint {http://arxiv.org/abs/1907.02903 }
		{arXiv:1907.02903  [hep-th]}\BibitemShut
	{NoStop}%
	\bibitem [{\citenamefont {Borchardt}\ and\ \citenamefont
		{Knorr}(2015)}]{Borchardt:2015rxa}%
	\BibitemOpen
	\bibfield  {author} {\bibinfo {author} {\bibfnamefont {Julia}\ \bibnamefont
			{Borchardt}}\ and\ \bibinfo {author} {\bibfnamefont {Benjamin}\ \bibnamefont
			{Knorr}},\ }\bibfield  {title} {\enquote {\bibinfo {title} {{Global solutions
					of functional fixed point equations via pseudospectral methods}},}\ }\href
	{\doibase 10.1103/PhysRevD.93.089904, 10.1103/PhysRevD.91.105011} {\bibfield
		{journal} {\bibinfo  {journal} {Phys. Rev.}\ }\textbf {\bibinfo {volume}
			{D91}},\ \bibinfo {pages} {105011} (\bibinfo {year} {2015})},\ \bibinfo
	{note} {[Erratum: Phys. Rev.D93,no.8,089904(2016)]},\ \Eprint
	{http://arxiv.org/abs/1502.07511} {arXiv:1502.07511 [hep-th]} \BibitemShut
	{NoStop}%
	\bibitem [{\citenamefont {Borchardt}\ and\ \citenamefont
		{Knorr}(2016)}]{Borchardt:2016pif}%
	\BibitemOpen
	\bibfield  {author} {\bibinfo {author} {\bibfnamefont {Julia}\ \bibnamefont
			{Borchardt}}\ and\ \bibinfo {author} {\bibfnamefont {Benjamin}\ \bibnamefont
			{Knorr}},\ }\bibfield  {title} {\enquote {\bibinfo {title} {{Solving
					functional flow equations with pseudo-spectral methods}},}\ }\href {\doibase
		10.1103/PhysRevD.94.025027} {\bibfield  {journal} {\bibinfo  {journal} {Phys.
				Rev.}\ }\textbf {\bibinfo {volume} {D94}},\ \bibinfo {pages} {025027}
		(\bibinfo {year} {2016})},\ \Eprint {http://arxiv.org/abs/1603.06726}
	{arXiv:1603.06726 [hep-th]} \BibitemShut {NoStop}%
	\bibitem [{\citenamefont {Boyd}(1987)}]{Boyd:ratCheb}%
	\BibitemOpen
	\bibfield  {author} {\bibinfo {author} {\bibfnamefont {J.P.}\ \bibnamefont
			{Boyd}},\ }\bibfield  {title} {\enquote {\bibinfo {title} {{Orthogonal
					rational functions on a semi-infinite interval}},}\ }\href@noop {} {\bibfield
		{journal} {\bibinfo  {journal} {J.Comp.Phys.}\ }\textbf {\bibinfo {volume}
			{70}},\ \bibinfo {pages} {63--88} (\bibinfo {year} {1987})}\BibitemShut
	{NoStop}%
	\bibitem [{\citenamefont {Boyd}(2000)}]{Boyd:ChebyFourier}%
	\BibitemOpen
	\bibfield  {author} {\bibinfo {author} {\bibfnamefont {John~P.}\ \bibnamefont
			{Boyd}},\ }\href@noop {} {\emph {\bibinfo {title} {{Chebyshev and Fourier
					Spectral Methods}}}},\ \bibinfo {edition} {2nd}\ ed.\ (\bibinfo  {publisher}
	{Dover Publications},\ \bibinfo {year} {2000})\BibitemShut {NoStop}%
	\bibitem [{\citenamefont {Reuter}\ and\ \citenamefont
		{Saueressig}(2002)}]{Reuter:2001ag}%
	\BibitemOpen
	\bibfield  {author} {\bibinfo {author} {\bibfnamefont {M.}~\bibnamefont
			{Reuter}}\ and\ \bibinfo {author} {\bibfnamefont {Frank}\ \bibnamefont
			{Saueressig}},\ }\bibfield  {title} {\enquote {\bibinfo {title}
			{{Renormalization group flow of quantum gravity in the Einstein-Hilbert
					truncation}},}\ }\href {\doibase 10.1103/PhysRevD.65.065016} {\bibfield
		{journal} {\bibinfo  {journal} {Phys. Rev.}\ }\textbf {\bibinfo {volume}
			{D65}},\ \bibinfo {pages} {065016} (\bibinfo {year} {2002})},\ \Eprint
	{http://arxiv.org/abs/hep-th/0110054} {arXiv:hep-th/0110054 [hep-th]}
	\BibitemShut {NoStop}%
	\bibitem [{\citenamefont {Reuter}\ and\ \citenamefont
		{Weyer}(2004)}]{Reuter:2004nx}%
	\BibitemOpen
	\bibfield  {author} {\bibinfo {author} {\bibfnamefont {M.}~\bibnamefont
			{Reuter}}\ and\ \bibinfo {author} {\bibfnamefont {H.}~\bibnamefont {Weyer}},\
	}\bibfield  {title} {\enquote {\bibinfo {title} {{Quantum gravity at
					astrophysical distances?}}}\ }\href {\doibase 10.1088/1475-7516/2004/12/001}
	{\bibfield  {journal} {\bibinfo  {journal} {JCAP}\ }\textbf {\bibinfo
			{volume} {0412}},\ \bibinfo {pages} {001} (\bibinfo {year} {2004})},\ \Eprint
	{http://arxiv.org/abs/hep-th/0410119} {arXiv:hep-th/0410119 [hep-th]}
	\BibitemShut {NoStop}%
	\bibitem [{\citenamefont {Saueressig}\ \emph {et~al.}(2019)\citenamefont
		{Saueressig}, \citenamefont {Gubitosi},\ and\ \citenamefont
		{Ripken}}]{Saueressig:2019fmx}%
	\BibitemOpen
	\bibfield  {author} {\bibinfo {author} {\bibfnamefont {Frank}\ \bibnamefont
			{Saueressig}}, \bibinfo {author} {\bibfnamefont {Giulia}\ \bibnamefont
			{Gubitosi}}, \ and\ \bibinfo {author} {\bibfnamefont {Chris}\ \bibnamefont
			{Ripken}},\ }\bibfield  {title} {\enquote {\bibinfo {title} {{Scales and
					hierachies in asymptotically safe quantum gravity: a review}},}\ }\href@noop
	{} {\  (\bibinfo {year} {2019})},\ \Eprint {http://arxiv.org/abs/1901.01731}
	{arXiv:1901.01731 [gr-qc]} \BibitemShut {NoStop}%
%
\bibitem{Giacchini:2016xns} 
B.~L.~Giacchini,
``On the cancellation of Newtonian singularities in higher-derivative gravity,''
Phys.\ Lett.\ B {\bf 766}, 306 (2017),
arXiv:1609.05432 [hep-th].
%
\bibitem{Calmet:2017qqa} 
X.~Calmet and B.~K.~El-Menoufi,
``Quantum Corrections to Schwarzschild Black Hole,''
Eur.\ Phys.\ J.\ C {\bf 77}, 243 (2017),
arXiv:1704.00261 [hep-th].
%
	\bibitem [{\citenamefont {Edholm}\ \emph {et~al.}(2016)\citenamefont {Edholm},
		\citenamefont {Koshelev},\ and\ \citenamefont {Mazumdar}}]{Edholm:2016hbt}%
	\BibitemOpen
	\bibfield  {author} {\bibinfo {author} {\bibfnamefont {James}\ \bibnamefont
			{Edholm}}, \bibinfo {author} {\bibfnamefont {Alexey~S.}\ \bibnamefont
			{Koshelev}}, \ and\ \bibinfo {author} {\bibfnamefont {Anupam}\ \bibnamefont
			{Mazumdar}},\ }\bibfield  {title} {\enquote {\bibinfo {title} {{Behavior of
					the Newtonian potential for ghost-free gravity and singularity-free
					gravity}},}\ }\href {\doibase 10.1103/PhysRevD.94.104033} {\bibfield
		{journal} {\bibinfo  {journal} {Phys. Rev.}\ }\textbf {\bibinfo {volume}
			{D94}},\ \bibinfo {pages} {104033} (\bibinfo {year} {2016})},\ \Eprint
	{http://arxiv.org/abs/1604.01989} {arXiv:1604.01989 [gr-qc]} \BibitemShut
	{NoStop}%
	\bibitem [{\citenamefont {Biswas}\ \emph {et~al.}(2006)\citenamefont {Biswas},
		\citenamefont {Mazumdar},\ and\ \citenamefont {Siegel}}]{Biswas:2005qr}%
	\BibitemOpen
	\bibfield  {author} {\bibinfo {author} {\bibfnamefont {Tirthabir}\
			\bibnamefont {Biswas}}, \bibinfo {author} {\bibfnamefont {Anupam}\
			\bibnamefont {Mazumdar}}, \ and\ \bibinfo {author} {\bibfnamefont {Warren}\
			\bibnamefont {Siegel}},\ }\bibfield  {title} {\enquote {\bibinfo {title}
			{{Bouncing universes in string-inspired gravity}},}\ }\href {\doibase
		10.1088/1475-7516/2006/03/009} {\bibfield  {journal} {\bibinfo  {journal}
			{JCAP}\ }\textbf {\bibinfo {volume} {0603}},\ \bibinfo {pages} {009}
		(\bibinfo {year} {2006})},\ \Eprint {http://arxiv.org/abs/hep-th/0508194}
	{arXiv:hep-th/0508194 [hep-th]} \BibitemShut {NoStop}%
	\bibitem [{\citenamefont {Modesto}(2012)}]{Modesto:2011kw}%
	\BibitemOpen
	\bibfield  {author} {\bibinfo {author} {\bibfnamefont {Leonardo}\
			\bibnamefont {Modesto}},\ }\bibfield  {title} {\enquote {\bibinfo {title}
			{{Super-renormalizable Quantum Gravity}},}\ }\href {\doibase
		10.1103/PhysRevD.86.044005} {\bibfield  {journal} {\bibinfo  {journal} {Phys.
				Rev.}\ }\textbf {\bibinfo {volume} {D86}},\ \bibinfo {pages} {044005}
		(\bibinfo {year} {2012})},\ \Eprint {http://arxiv.org/abs/1107.2403}
	{arXiv:1107.2403 [hep-th]} \BibitemShut {NoStop}%
	\bibitem [{\citenamefont {Biswas}\ \emph {et~al.}(2012)\citenamefont {Biswas},
		\citenamefont {Gerwick}, \citenamefont {Koivisto},\ and\ \citenamefont
		{Mazumdar}}]{Biswas:2011ar}%
	\BibitemOpen
	\bibfield  {author} {\bibinfo {author} {\bibfnamefont {Tirthabir}\
			\bibnamefont {Biswas}}, \bibinfo {author} {\bibfnamefont {Erik}\ \bibnamefont
			{Gerwick}}, \bibinfo {author} {\bibfnamefont {Tomi}\ \bibnamefont
			{Koivisto}}, \ and\ \bibinfo {author} {\bibfnamefont {Anupam}\ \bibnamefont
			{Mazumdar}},\ }\bibfield  {title} {\enquote {\bibinfo {title} {{Towards
					singularity and ghost free theories of gravity}},}\ }\href {\doibase
		10.1103/PhysRevLett.108.031101} {\bibfield  {journal} {\bibinfo  {journal}
			{Phys. Rev. Lett.}\ }\textbf {\bibinfo {volume} {108}},\ \bibinfo {pages}
		{031101} (\bibinfo {year} {2012})},\ \Eprint {http://arxiv.org/abs/1110.5249}
	{arXiv:1110.5249 [gr-qc]} \BibitemShut {NoStop}%
	\bibitem [{\citenamefont {Modesto}\ and\ \citenamefont
		{Rachwal}(2017)}]{Modesto:2017sdr}%
	\BibitemOpen
	\bibfield  {author} {\bibinfo {author} {\bibfnamefont {Leonardo}\
			\bibnamefont {Modesto}}\ and\ \bibinfo {author} {\bibfnamefont {Leslaw}\
			\bibnamefont {Rachwal}},\ }\bibfield  {title} {\enquote {\bibinfo {title}
			{{Nonlocal quantum gravity: A review}},}\ }\href {\doibase
		10.1142/S0218271817300208} {\bibfield  {journal} {\bibinfo  {journal} {Int.
				J. Mod. Phys.}\ }\textbf {\bibinfo {volume} {D26}},\ \bibinfo {pages}
		{1730020} (\bibinfo {year} {2017})}\BibitemShut {NoStop}%
	\bibitem [{\citenamefont {Connes}(1996)}]{Connes:1996gi}%
	\BibitemOpen
	\bibfield  {author} {\bibinfo {author} {\bibfnamefont {Alain}\ \bibnamefont
			{Connes}},\ }\bibfield  {title} {\enquote {\bibinfo {title} {{Gravity coupled
					with matter and foundation of noncommutative geometry}},}\ }\href {\doibase
		10.1007/BF02506388} {\bibfield  {journal} {\bibinfo  {journal} {Commun. Math.
				Phys.}\ }\textbf {\bibinfo {volume} {182}},\ \bibinfo {pages} {155--176}
		(\bibinfo {year} {1996})},\ \Eprint {http://arxiv.org/abs/hep-th/9603053}
	{arXiv:hep-th/9603053 [hep-th]} \BibitemShut {NoStop}%
	\bibitem [{\citenamefont {van~den Dungen}\ and\ \citenamefont {van
			Suijlekom}(2012)}]{vandenDungen:2012ky}%
	\BibitemOpen
	\bibfield  {author} {\bibinfo {author} {\bibfnamefont {Koen}\ \bibnamefont
			{van~den Dungen}}\ and\ \bibinfo {author} {\bibfnamefont {Walter~D.}\
			\bibnamefont {van Suijlekom}},\ }\bibfield  {title} {\enquote {\bibinfo
			{title} {{Particle Physics from Almost Commutative Spacetimes}},}\ }\href
	{\doibase 10.1142/S0129055X1230004X} {\bibfield  {journal} {\bibinfo
			{journal} {Rev. Math. Phys.}\ }\textbf {\bibinfo {volume} {24}},\ \bibinfo
		{pages} {1230004} (\bibinfo {year} {2012})},\ \Eprint
	{http://arxiv.org/abs/1204.0328} {arXiv:1204.0328 [hep-th]} \BibitemShut
	{NoStop}%
	\bibitem [{\citenamefont {Connes}\ and\ \citenamefont {Marcolli}()}]{1414300}%
	\BibitemOpen
	\bibfield  {author} {\bibinfo {author} {\bibfnamefont {Alain}\ \bibnamefont
			{Connes}}\ and\ \bibinfo {author} {\bibfnamefont {Matilde}\ \bibnamefont
			{Marcolli}},\ }\href {http://www.alainconnes.org/docs/bookwebfinal.pdf}
	{\emph {\bibinfo {title} {{Noncommutative Geometry, Quantum Fields and
					Motives}}}}\BibitemShut {NoStop}%
	\bibitem [{\citenamefont {Kurkov}\ \emph {et~al.}(2014)\citenamefont {Kurkov},
		\citenamefont {Lizzi},\ and\ \citenamefont {Vassilevich}}]{Kurkov:2013kfa}%
	\BibitemOpen
	\bibfield  {author} {\bibinfo {author} {\bibfnamefont {M.~A.}\ \bibnamefont
			{Kurkov}}, \bibinfo {author} {\bibfnamefont {Fedele}\ \bibnamefont {Lizzi}},
		\ and\ \bibinfo {author} {\bibfnamefont {Dmitri}\ \bibnamefont
			{Vassilevich}},\ }\bibfield  {title} {\enquote {\bibinfo {title} {{High
					energy bosons do not propagate}},}\ }\href {\doibase
		10.1016/j.physletb.2014.02.053} {\bibfield  {journal} {\bibinfo  {journal}
			{Phys. Lett.}\ }\textbf {\bibinfo {volume} {B731}},\ \bibinfo {pages}
		{311--315} (\bibinfo {year} {2014})},\ \Eprint
	{http://arxiv.org/abs/1312.2235} {arXiv:1312.2235 [hep-th]} \BibitemShut
	{NoStop}%
	\bibitem [{\citenamefont {Alkofer}\ \emph {et~al.}(2015)\citenamefont
		{Alkofer}, \citenamefont {Saueressig},\ and\ \citenamefont
		{Zanusso}}]{Alkofer:2014raa}%
	\BibitemOpen
	\bibfield  {author} {\bibinfo {author} {\bibfnamefont {Natalia}\ \bibnamefont
			{Alkofer}}, \bibinfo {author} {\bibfnamefont {Frank}\ \bibnamefont
			{Saueressig}}, \ and\ \bibinfo {author} {\bibfnamefont {Omar}\ \bibnamefont
			{Zanusso}},\ }\bibfield  {title} {\enquote {\bibinfo {title} {{Spectral
					dimensions from the spectral action}},}\ }\href {\doibase
		10.1103/PhysRevD.91.025025} {\bibfield  {journal} {\bibinfo  {journal} {Phys.
				Rev.}\ }\textbf {\bibinfo {volume} {D91}},\ \bibinfo {pages} {025025}
		(\bibinfo {year} {2015})},\ \Eprint {http://arxiv.org/abs/1410.7999}
	{arXiv:1410.7999 [hep-th]} \BibitemShut {NoStop}%
\end{thebibliography}
\end{document}